\documentclass[ALICE,manyauthors]{cernphprep}
\usepackage[comma,square,numbers,sort&compress]{natbib}
\usepackage{amsmath}
\usepackage{hyperref}
\usepackage{lineno}
\usepackage{xspace}
\usepackage{comment}
\usepackage{rotating}
\usepackage[utf8]{inputenc}
\usepackage{amsmath}
\usepackage[normalem]{ulem}
\usepackage[T1]{fontenc}
\usepackage{hepnames}
\usepackage{pennames}
\usepackage{xcolor}

\begin{document}
%

\newcommand{\pp}           {pp\xspace}
\newcommand{\ppbar}        {\mbox{$\mathrm {p\overline{p}}$}\xspace}
\newcommand{\XeXe}         {\mbox{Xe--Xe}\xspace}
\newcommand{\PbPb}         {\mbox{Pb--Pb}\xspace}
\newcommand{\pA}           {\mbox{pA}\xspace}
\newcommand{\pPb}          {\mbox{p--Pb}\xspace}
\newcommand{\AuAu}         {\mbox{Au--Au}\xspace}
\newcommand{\dAu}          {\mbox{d--Au}\xspace}

\newcommand{\s}            {\ensuremath{\sqrt{s}}\xspace}
\newcommand{\snn}          {\ensuremath{\sqrt{s_{\mathrm{NN}}}}\xspace}
\newcommand{\pt}           {\ensuremath{p_{\rm T}}\xspace}
\newcommand{\meanpt}       {$\langle p_{\mathrm{T}}\rangle$\xspace}
\newcommand{\ycms}         {\ensuremath{y_{\rm CMS}}\xspace}
\newcommand{\ylab}         {\ensuremath{y_{\rm lab}}\xspace}
\newcommand{\etarange}[1]  {\mbox{$\left | \eta \right |~<~#1$}}
\newcommand{\yrange}[1]    {\mbox{$\left | y \right |~<~#1$}}
\newcommand{\dndy}         {\ensuremath{\mathrm{d}N_\mathrm{ch}/\mathrm{d}y}\xspace}
\newcommand{\dndeta}       {\ensuremath{\mathrm{d}N_\mathrm{ch}/\mathrm{d}\eta}\xspace}
\newcommand{\avdndeta}     {\ensuremath{\langle\dndeta\rangle}\xspace}
\newcommand{\dNdy}         {\ensuremath{\mathrm{d}N_\mathrm{ch}/\mathrm{d}y}\xspace}
\newcommand{\Npart}        {\ensuremath{N_\mathrm{part}}\xspace}
\newcommand{\Ncoll}        {\ensuremath{N_\mathrm{coll}}\xspace}
\newcommand{\dEdx}         {\ensuremath{\textrm{d}E/\textrm{d}x}\xspace}
\newcommand{\RpPb}         {\ensuremath{R_{\rm pPb}}\xspace}

\newcommand{\nineH}        {$\sqrt{s}~=~0.9$~Te\kern-.1emV\xspace}
\newcommand{\seven}        {$\sqrt{s}~=~7$~Te\kern-.1emV\xspace}
\newcommand{\twoH}         {$\sqrt{s}~=~0.2$~Te\kern-.1emV\xspace}
\newcommand{\twosevensix}  {$\sqrt{s}~=~2.76$~Te\kern-.1emV\xspace}
\newcommand{\five}         {$\sqrt{s}~=~5.02$~Te\kern-.1emV\xspace}
\newcommand{\twosevensixnn}{$\sqrt{s_{\mathrm{NN}}}~=~2.76$~Te\kern-.1emV\xspace}
\newcommand{\fivenn}       {$\sqrt{s_{\mathrm{NN}}}~=~5.02$~Te\kern-.1emV\xspace}
\newcommand{\LT}           {L{\'e}vy-Tsallis\xspace}
\newcommand{\GeVc}         {Ge\kern-.1emV/$c$\xspace}
\newcommand{\MeVc}         {Me\kern-.1emV/$c$\xspace}
\newcommand{\TeV}          {Te\kern-.1emV\xspace}
\newcommand{\GeV}          {Ge\kern-.1emV\xspace}
\newcommand{\MeV}          {Me\kern-.1emV\xspace}
\newcommand{\GeVmass}      {Ge\kern-.2emV/$c^2$\xspace}
\newcommand{\MeVmass}      {Me\kern-.2emV/$c^2$\xspace}
\newcommand{\lumi}         {\ensuremath{\mathcal{L}}\xspace}

\newcommand{\ITS}          {\rm{ITS}\xspace}
\newcommand{\TOF}          {\rm{TOF}\xspace}
\newcommand{\ZDC}          {\rm{ZDC}\xspace}
\newcommand{\ZDCs}         {\rm{ZDCs}\xspace}
\newcommand{\ZNA}          {\rm{ZNA}\xspace}
\newcommand{\ZNC}          {\rm{ZNC}\xspace}
\newcommand{\SPD}          {\rm{SPD}\xspace}
\newcommand{\SDD}          {\rm{SDD}\xspace}
\newcommand{\SSD}          {\rm{SSD}\xspace}
\newcommand{\TPC}          {\rm{TPC}\xspace}
\newcommand{\TRD}          {\rm{TRD}\xspace}
\newcommand{\VZERO}        {\rm{V0}\xspace}
\newcommand{\VZEROA}       {\rm{V0A}\xspace}
\newcommand{\VZEROC}       {\rm{V0C}\xspace}
\newcommand{\Vdecay} 	   {\ensuremath{V^{0}}\xspace}

\newcommand{\ee}           {\ensuremath{\rm e^{+}e^{-}}} 
\newcommand{\pip}          {\ensuremath{\pi^{+}}\xspace}
\newcommand{\pim}          {\ensuremath{\pi^{-}}\xspace}
\newcommand{\kap}          {\ensuremath{\rm{K}^{+}}\xspace}
\newcommand{\kam}          {\ensuremath{\rm{K}^{-}}\xspace}
\newcommand{\pbar}         {\ensuremath{\rm\overline{p}}\xspace}
\newcommand{\kzero}        {\ensuremath{{\rm K}^{0}_{\rm{S}}}\xspace}
\newcommand{\lmb}          {\ensuremath{\Lambda}\xspace}
\newcommand{\almb}         {\ensuremath{\overline{\Lambda}}\xspace}
\newcommand{\Om}           {\ensuremath{\Omega^-}\xspace}
\newcommand{\Mo}           {\ensuremath{\overline{\Omega}^+}\xspace}
\newcommand{\X}            {\ensuremath{\Xi^-}\xspace}
\newcommand{\Ix}           {\ensuremath{\overline{\Xi}^+}\xspace}
\newcommand{\Xis}          {\ensuremath{\Xi^{\pm}}\xspace}
\newcommand{\Oms}          {\ensuremath{\Omega^{\pm}}\xspace}
\newcommand{\degree}       {\ensuremath{^{\rm o}}\xspace}
\newcommand{\PJpsi}         {\ensuremath{\rm{J}/\psi}\xspace}
\newcommand {\lumint} {{\cal L}_{\rm int}}
\newcommand{\fb}       {\ensuremath{f_{\rm b}}\xspace}

\newcommand{\APbottom}{\ensuremath{\rm {\overline{b}}}\xspace}
\newcommand{\Pbottom}{\ensuremath{\rm {b}}\xspace}
\newcommand{\APproton}{\ensuremath{\rm {\overline{p}}}\xspace}
\newcommand{\Pcgc}{\ensuremath{\chi_{\rm{c}}}\xspace}

\begin{titlepage}
\PHyear{2021}       
\PHnumber{077}      
\PHdate{7 May}  

\title{Inclusive, prompt and non-prompt \PJpsi\ production at midrapidity in p$-$Pb collisions at $\snn = 5.02$ \TeV}
\ShortTitle{\PJpsi production in p$-$Pb collisions at midrapidity at $\snn = 5.02$ \TeV}   

\Collaboration{ALICE Collaboration\thanks{See Appendix~\ref{app:collab} for the list of collaboration members}}
\ShortAuthor{ALICE Collaboration} 

\begin{abstract}
A measurement of inclusive, prompt, and non-prompt \PJpsi\ production in p$-$Pb collisions at a nucleon--nucleon centre-of-mass energy $\snn = 5.02$ \TeV is presented. The inclusive \PJpsi mesons are reconstructed in the dielectron decay channel at midrapidity down to a transverse momentum $\pt = 0$. 
The 
inclusive \PJpsi 
nuclear modification factor \RpPb 
is calculated by comparing the new results in p$-$Pb collisions  
to a recently 
measured proton$-$proton reference at the same centre-of-mass energy. 
Non-prompt \PJpsi mesons, which originate from the decay of beauty hadrons, are separated from promptly produced \PJpsi on a statistical basis for $\pt$ larger than 1.0 GeV/$c$. 
These results are based on the data sample collected by the ALICE detector during the 2016 LHC p$-$Pb run, corresponding to an integrated luminosity 
$\lumint = 292 \pm 11 \; {\rm \mu b}^{-1}$, which is 
six times larger than 
the 
previous publications. 
The total uncertainty on the \pt-integrated inclusive \PJpsi and non-prompt \PJpsi cross section are reduced by a factor 1.7 and 2.2, respectively. 
The 
measured cross sections and \RpPb 
are compared with
theoretical models that include 
various combinations of cold nuclear matter effects. 
From the non-prompt \PJpsi production cross section, the $\Pbottom\APbottom$ production cross section at midrapidity, $\mathrm{d}\sigma_{\Pbottom\APbottom}/\mathrm{d} y$, and the total cross section extrapolated over full phase space, $\sigma_{\rm b\overline{b}}$, are derived. 

\end{abstract}
\end{titlepage}

\setcounter{page}{2} 


\section{Introduction} 

The production of the \PJpsi meson in hadronic interactions represents a challenging testing ground for models based on quantum chromodynamics (QCD). 
In proton$-$proton (pp) and proton$-$antiproton (p\APproton) collisions, charmonium production has been intensively studied experimentally at the Tevatron~\cite{D0_1,D0_2,CDF1,CDF2}, 
RHIC~\cite{Adare:2006kf,Adam:2019mrg} and the LHC~\cite{OurppRefPaper,Abelev:2012kr,Aamodt:2011gj,Adam:2016rdg,Adam:2015rta,Acharya:2017hjh,Abelev:2012gx,Aaboud:2017cif,Sirunyan:2017mzd,Aaij:2012asz,Aaij:2011jh,Aaij:2013yaa,Aaij:2015rla}, and it can be 
described in the framework of the non-relativistic quantum chromodynamics (NRQCD) effective theory~\cite{NRQCD1,NRQCD2,NRQCD3,BRAMBILLA,SAPOREGRAVIS,ImprovedNRQCD2,Han:2015eta,Zhang:2015eta,Butenschoen:2015eta}. 
    
In heavy-ion collisions, charmonium production is highly sensitive to the nature of the hot and dense matter created in these collisions, the quark$-$gluon plasma (QGP), see Ref.~\cite{BRAMBILLA,SAPOREGRAVIS,LEIDEN} for recent reviews. For a precise interpretation of the heavy-ion results, detailed comparisons with both the reference results obtained in elementary pp collisions and those in proton$-$nucleus (p$-$A) collisions are indispensable. 
The latter is used to disentangle effects
due to interaction between the charmonium states and the QGP medium created in heavy-ion collisions from those that can be ascribed to cold nuclear matter (CNM).
In fact, 
the nuclear environment affects the free nucleon parton distribution functions (PDFs), inducing modifications that depend on the parton fractional momentum $x_{\rm B}$, the four-momentum transfer squared ($Q^2$) and the mass number A, as first discovered by the European Muon Collaboration~\cite{EMC}. The modified distributions can be described using nuclear parton distribution functions (nPDFs)~\cite{Kusina:2015vfa,Eskola:2016oht,AbdulKhalek:2020yuc}.
In the $x_{\rm B}$ and $Q^2$ domain of ``shadowing'' reached for nuclear collisions at LHC energies in 
charm and beauty 
production, the parton density, and most notably the one of the gluons, is reduced with respect to the free nucleon~\cite{Eskola:2019bgf,Eskola:2020yfa,Kusina:2020dki}.
At very small $x_{\rm B}$ values, where the gluon density becomes very large, the nuclear environment is expected to favour a saturation process, which can be described using the colour glass condensate (CGC) effective theory ~\cite{CGC,CGCreview,Fujii:2013yja}. 
In addition, in the nuclear environment
partons can lose energy via initial-state radiation, thus 
reducing 
the centre-of-mass
energy of the partonic system~\cite{Vitev}, experience transverse
momentum broadening due to multiple soft collisions before
the production of the $\Pcharm\APcharm$ pair~\cite{Lev,Wang,Kopeliovich}, or loose energy through coherent effects~\cite{Arleo:2013zua}. 
Finally, 
once produced, the charmonium state
could be dissociated via inelastic interactions with the surrounding nucleons~\cite{Gerschel}. This process, which is 
dominant among the CNM effects at low collision energy~\cite{NA50,PHENIX}, 
should become negligible at the LHC, where the crossing time of the two nuclei is much shorter 
than the resonance formation time~\cite{Hufner,Kharzeev,ALICEPsi2S}.

The comparison of the production measured in p$-$A collisions to the one in pp collisions allows the CNM effects to be constrained. The size of these effects can be quantified by the nuclear modification factor, which is defined as the production cross section in p$-$A collisions ($\sigma_{\rm pA}$) divided by that in pp collisions ($\sigma_{\rm pp})$ scaled by the mass number $A$, 
\begin{equation}
	R_{\rm pA}(y,p_{\rm T}) = 
    \frac{1}{\rm A}
	\frac{{\rm d}^2\sigma_{\rm pA} / {\rm d}y {\rm d}p_{\rm T}}
             {{\rm d}^2\sigma_{\rm pp} / {\rm d}y {\rm d}p_{\rm T}}  ,
\label{RpA}
\end{equation}
where $y$ is the rapidity of the observed hadron in the nucleon$-$nucleon centre-of-mass frame, and \pt\ its transverse momentum.   
In the absence of nuclear effects, $R_{\rm pA}$ is expected to be equal to unity. 

The inclusive \PJpsi yield 
is composed of three contributions: prompt \PJpsi produced directly in the primary hadronic collision, prompt \PJpsi produced indirectly via the decay of heavier charmonium states such as \Pcgc and \PpsiTwoS, and non-prompt \PJpsi from the decay of beauty hadrons ($\rm h_b$). 
By subtracting the non-prompt component from the inclusive \PJpsi production, more direct comparisons with models describing the charmonium production can be considered. However, this contribution is not large, in particular at low \pt 
($\approx 10$--$20$\% for $\pt < 10 $~GeV/$c$~\cite{OurpPbPaper}), where the bulk of the production is located, and the inclusive \PJpsi production represents already a valuable observable.   

The measurement of the non-prompt component gives access to study open beauty production through the inclusive decay channel $\rm h_b \rightarrow \PJpsi + X $. 
In pp collisions, the production cross section
of beauty hadrons can be computed with factorisation approaches, either in terms of $Q^2$ (collinear factorisation) ~\cite{Factorisation1}  as a convolution of the PDFs of the incoming protons, the partonic hard-scattering cross sections, and the fragmentation functions, or of the partonic transverse momentum $k_{\rm T}$~\cite{Factorisation2}. 
The cold-medium processes that affect the charmonium production in proton$-$nucleus and nucleus$-$nucleus collisions can also affect the beauty hadron production~\cite{Kusina:2015vfa,Eskola:2016oht,AbdulKhalek:2020yuc,Eskola:2019bgf,Eskola:2020yfa,Kusina:2020dki,CGC,CGCreview,Fujii:2013yja,Lev,Wang,Kopeliovich,Vitev,Arleo:2010rb}. 
Also in this case, the nuclear modification factor $R_{\rm pA}$ can be useful to study these effects.

Charmonium and open-beauty production cross sections in p$-$Pb collisions have been measured at LHC energies by the ALICE~\cite{ALICEJPSI,ALICEPsi2S,Adam:2015jsa,Adam:2015iga,OurpPbPaper,Acharya:2018kxc,Adam:2016wyz}, ATLAS~\cite{Aad:2015ddl,Aaboud:2017cif},
CMS~\cite{Khachatryan:2015sva,Khachatryan:2015uja,Sirunyan:2017mzd,Sirunyan:2018pse}
and LHCb~\cite{Aaij:2013zxa,Aaij:2016eyl,Aaij:2017cqq,Aaij:2019lkm} collaborations over a wide range of rapidity and transverse momentum. Thanks to its moderate magnetic field, particle identification capability, and low material budget of the tracking system in the central barrel, the ALICE apparatus has a unique coverage for \PJpsi measurements at midrapidity and low transverse momentum. Previous ALICE measurements were published based on the p$-$Pb data sample collected in 2013~\cite{Adam:2015iga,OurpPbPaper,Adam:2015jsa,Adamova:2017uhu}. This paper presents new measurements of the \pt-differential cross sections for the inclusive, prompt, and non-prompt \PJpsi production in p$-$Pb collisions at $\snn=5.02$~TeV, using the data sample collected in 2016, which is six times larger than that of 2013. Moreover, the cross section of inclusive \PJpsi production measured in pp collisions at $\sqrt{s}=5.02$~TeV~\cite{OurppRefPaper} is used to derive the $R_{\rm pA}$ results instead of the interpolation procedure adopted  in the previous p–Pb publication~\cite{Adam:2015iga}. 
Therefore, the new results, which  are significantly more precise 
and are obtained differentially in \pt and in finer \pt intervals, 
supersede the measurements published in Refs.~\cite{Adam:2015iga,OurpPbPaper}.

\section{Data Analysis}

A complete description of the ALICE apparatus and its performance is presented in Refs.~\cite{Aamodt:2008zz,Abelev:2014ffa}. The central-barrel detectors employed for the analysis presented in this paper are the Inner Tracking System (\ITS) and the Time Projection Chamber (\TPC). The \ITS~\cite{ITS} provides tracking and vertex reconstruction close to the interaction point (IP). It is made up of six concentric cylindrical layers of silicon detectors surrounding the beam pipe with radial positions between 3.9 cm and 43.0 cm. The two innermost layers consist of Silicon Pixel Detectors (\SPD), the two central layers are made up of Silicon Drift Detectors (\SDD), and the two outermost layers of Silicon Strip Detectors (\SSD).
The \TPC~\cite{TPC} consists of a large cylindrical drift chamber surrounding the \ITS and extending from 85 cm to 247 cm along the radial direction and from $-$250 cm to $+$250 cm along the beam direction ($z$) relative to the IP. 
It is the main ALICE tracking device and allows also charged particles to be identified through specific energy loss ($\textup{d}E/\textup{d}x$) measurements in the detector gas. 
Both the \TPC and \ITS are embedded in a solenoidal magnet that generates a 0.5 T magnetic field along the beam direction. 
They cover the pseudorapidity interval $|\eta|<$ 0.9 and allow \PJpsi mesons to be reconstructed through the $\rm e^+e^-$ decay channel in the central rapidity region down to zero \pt.

The measurements presented in this paper are based on the set of minimum bias (MB) p$-$Pb collisions at $\snn = 5.02$ \TeV collected in 2016 during the LHC Run 2 data taking period, corresponding to an integrated luminosity $\lumint = 292 \pm 11\  {\rm \mu b}^{-1}$. The latter is determined from the number of MB events and the MB-trigger cross section, which was measured via a van der Meer scan, with negligible statistical uncertainty and a systematic uncertainty of 3.7\%~\cite{Abelev:2014epa}.
Collisions were realized by delivering proton and Pb beams with energies of 4 \TeV and 1.58 \TeV per nucleon, respectively. 
The proton and Pb beams circulated in the LHC anticlockwise and clockwise, respectively, during the period of data taking considered for this analysis.  
The MB trigger condition is provided by the \VZERO detector~\cite{VZERO}: a system made up of two arrays of plastic scintillators placed 
on either side of the IP and 
covering the full azimuthal angle and the pseudorapidity intervals $2.8 < \eta < 5.1$ and $-3.7 < \eta < -1.7$. 
The trigger condition required at least one hit in both 
the two arrays 
during the nominal bunch crossing time frame, allowing non-single-diffractive p$-$Pb collisions to be selected with an efficiency higher than 99\%~\cite{ALICE:2012xs}. The timing information from the \VZERO detectors is also used, in combination with that from the \SPD, to implement an offline rejection of beam-induced background interactions occurring outside the nominal colliding bunch crossings. Events with more than one interaction per bunch crossing are reduced down to a negligible amount by means of a dedicated algorithm employing reconstructed tracks to detect the presence of  multiple collision vertices. Only collision events with a reconstructed primary vertex lying within $\pm 10$~cm from the nominal IP along the beam direction are considered in order to obtain a uniform coverage for the central-barrel detectors.
An event sample of about $6\times 10^8$ MB events is obtained after the application of the above described selection criteria.

\subsection{Inclusive \PJpsi}
\label{sec:inclusive}

\begin{figure}[b]
    \begin{center}
    \includegraphics[width = 0.98\textwidth]{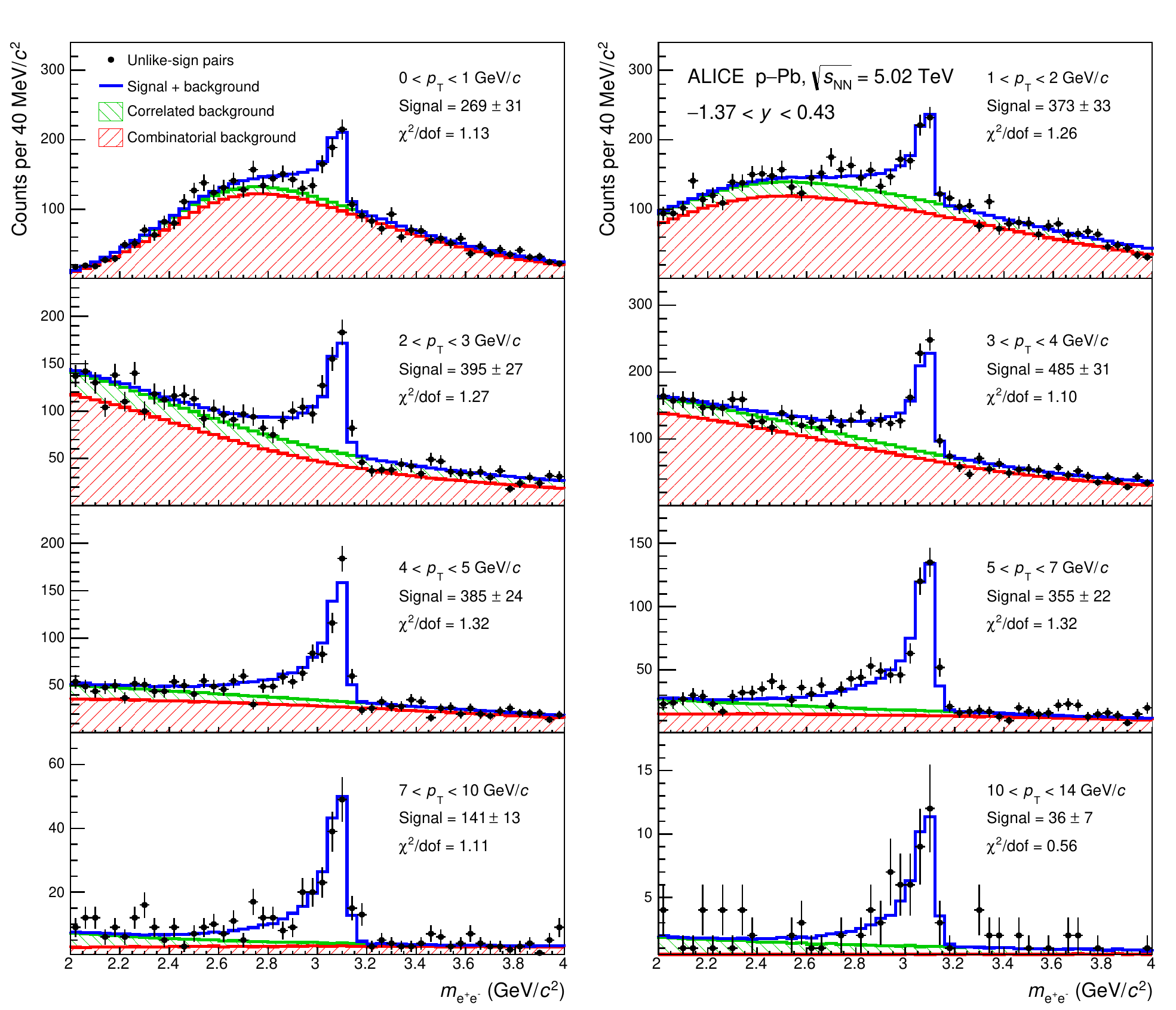}
    \end{center}
    \caption{Opposite-sign dielectron invariant mass distributions for the \pt-intervals used for this analysis. The signal plus total background (blue), the combinatorial background (red), and the correlated background (green), evaluated as described in the text, are shown separately in each panel.
    The $\chi^2/{\rm ndf}$ values of the signal template plus the total background function are also reported along with the raw yields in the range 
    $2.92~<~m_{\rm e^+e^-}~<~3.16$~GeV/$c^2$.} 
    \label{fig:technincalInclusive}
\end{figure}

Electron candidates are selected following similar procedures as those described in Ref.~\cite{Adam:2015iga}. 
The tracks, reconstructed with the \ITS and \TPC detectors, are required to have a transverse momentum $\pt^{\rm e}~>~1.0$~GeV/$c$ and pseudorapidity $|\eta^{\rm e}| < 0.9$, as well as at least 70 (out of a maximum of 159) attached TPC clusters and a track fit $\chi^2/{\rm dof} < 2$ in order to ensure a uniform tracking efficiency in the \TPC. 
Electron identification is performed by requiring the measured $\textup{d}E/\textup{d}x$ to be compatible with the expected specific energy loss for electrons within $3 \sigma$, with $\sigma$ denoting the specific energy-loss resolution of the \TPC. 
Tracks compatible with the pion and proton energy loss expectations within $3\sigma$ are rejected.
At least one hit in either of the two \SPD layers is required to remove background electrons produced from the conversion of photons in the detector materials at large radii.
Additional suppression of this background 
is realized by discarding electron (positron) candidates, which are compatible with a photon conversion when combined with a positron (electron) candidate of the same event, 
through the application of dedicated topological selections.
These selections were verified, employing Monte Carlo (MC) simulations, to have a negligible impact on the \PJpsi signal.
Finally, in order to reduce the overall background at low transverse momentum, a set of slightly tighter \SPD and particle identification (PID) requirements is applied to electrons and positrons forming candidate pairs with $\pt < 3$ GeV/$c$. A hit in the first \SPD layer and a $3.5\sigma$ pion and proton rejection condition is required instead of $3 \sigma$ 
for higher \pt values.
The sample of \PJpsi candidates is obtained by combining the selected opposite-sign tracks in the same event and requiring the \PJpsi rapidity to be within $|y_{\rm lab} | < 0.9$ in the laboratory system. 
Due to the energy asymmetry of the proton and lead beams, such a requirement corresponds to a selection of \PJpsi candidates within $-1.37 < y < 0.43$ in the nucleon$-$nucleon centre-of-mass system.
The resulting dielectron invariant mass ($m_{\rm e^+e^-}$) distributions are shown in Fig.~\ref{fig:technincalInclusive} for eight selected transverse momentum intervals, from 0 to 14 GeV/$c$. 
The signal component is characterised by an asymmetric shape, with a long tail towards low invariant masses due to the \PJpsi radiative decay channel ($\PJpsi \rightarrow \ee \gamma$ ) and the bremsstrahlung-induced energy loss of daughter electrons in the detector material.
The background component is composed of both a combinatorial and a correlated part, with the latter mainly originating from the semileptonic decay of correlated open heavy-flavour hadrons.

The inclusive \PJpsi yield is determined from the invariant mass distributions 
using the same technique as described in 
Ref.~\cite{OurppRefPaper}. 
At first, the combinatorial background shape is modelled by means of a mixed event (ME) technique and then scaled to the invariant mass distribution of like-sign track pairs.
Then,  
the combinatorial background is subtracted from the opposite-sign dielectron invariant mass distribution,  and the correlated background is evaluated by fitting the resulting distribution with a two-component function composed of a MC template for the \PJpsi signal and of an empirical function for the correlated background. 
The latter is defined to be either an exponential or a combination of an exponential and a polynomial. The former is obtained by a detailed MC simulation of \PJpsi decays in the ALICE detectors, based on GEANT3~\cite{GEANT3} and the full reconstruction chain as for real events, which is then also used to correct the raw yield for the selection procedure and detector inefficiencies and described in details in the next paragraph.  
After subtracting from the opposite-sign dielectron invariant mass distribution also the correlated background, the raw \PJpsi yield is obtained 
by counting the number of entries within the invariant mass interval $2.92~<~m_{\rm e^+e^-}~<~3.16$~GeV/$c^2$. 
In Fig.~\ref{fig:technincalInclusive}, the different components used in the procedure to describe the opposite-sign dielectron invariant mass distributions are shown superimposed. 
An alternative method was also considered, where the invariant mass distribution after the subtraction of the correlated background is fitted with a Crystal Ball (CB) function~\cite{CrystalBall} for the signal plus either an exponential or a combination of an exponential and a polynomial for the background, and the raw yield is obtained from the integral of the  best-fit CB function. 
The alternative method yields 
results compatible with those from the standard approach.

In order to correct the raw yield 
for the chosen selection procedure as well as for detector inefficiencies, a MC simulation was implemented by injecting \PJpsi signal events into MB p$-$Pb collision events simulated with the EPOS-LHC model~\cite{EPOS-LHC}. 
The 
\PJpsi component was generated starting with \pt and $y$ distributions that match well a next-to-leading order (NLO) Colour Evaporation Model (CEM) calculations~\cite{CEM_0,CEM_1} with the inclusion of nuclear effects based on the EPS09 parameterisation~\cite{EPS09}. 
The \PJpsi decay into dielectrons was simulated using the EvtGen package~\cite{EVTGEN} in combination with the PHOTOS model~\cite{PHOTOS} in order to provide a proper description of the radiative decay channel. 
In the simulation, 
GEANT3~\cite{GEANT3} was used to reproduce the propagation of particles through the ALICE experimental setup, taking into account the response of the detectors. 
The same reconstruction procedure used for data was then applied to the simulated events in order to evaluate the product of acceptance times efficiency ($ A\times \epsilon$),
which accounts for: the detector acceptance, the track quality requirements, the electron identification criteria, and the fraction of the signal counted within the invariant mass interval $2.92 < m_{\rm e^+e^-} < 3.16$ GeV/$c^2$.
The $A\times \epsilon$ retrieved from MC exhibits a smooth and mild variation with the \PJpsi \pt, ranging from $\sim$8.5\% to $\sim$16\% in the \pt range from 0 to 14 GeV/$c$. 
Consequently, the resulting $\left\langle A\times \epsilon \right\rangle$ correction factor, which is the average of $A\times \epsilon$  over \pt in a finite-size \pt interval, shows a weak dependence on the \pt shape assumed in the simulation for the \PJpsi component. In the end, the final correction factors were computed by re-weighting the original MC distribution to best-fit the inclusive \PJpsi spectrum already measured in p--Pb collisions at the same energy~\cite{Adam:2015iga}. 

The $p_{\rm T}$-differential cross section for inclusive \PJpsi production is calculated as
 
\begin{equation}
    \frac{{\rm d}^2\sigma_{\PJpsi}} {{\rm d}y{\rm d}p_{\rm T}} = \frac{N_{\PJpsi}\left({\rm \Delta}y,{\rm \Delta}p_{\rm T}\right)} { {\rm BR}\left( \PJpsi \rightarrow \ee \right)\times \left\langle A\times \epsilon \right\rangle \left({\rm \Delta}y,{\rm \Delta}p_{\rm T}\right) \times {\rm \Delta}y \times {\rm \Delta}p_{\rm T} \times \lumint }\ ,
\end{equation}

where ${\rm \Delta}y = 1.8$ corresponds to the width of the analysed rapidity interval, ${\rm \Delta}p_{\rm T}$ is the width of the considered $p_{\rm T}$ interval,   $N_{\PJpsi}$ is the 
raw \PJpsi yield in the interval, 
and ${\rm BR}\left( \PJpsi \rightarrow \ee \right) = (5.97 \pm 0.03)\%$ is the branching ratio for \PJpsi decaying into dielectrons~\cite{PDG2018}. 

The inclusive \PJpsi nuclear modification factor is obtained, according to Eq.~\ref{RpA}, by  dividing the $\pt$-differential cross section by the reference cross section measured up to $\pt = 10$~GeV/$c$ in pp collisions at $\sqrt{s} = 5.02$ TeV~\cite{OurppRefPaper}. The rapidity shift of $\Delta y = 0.465$ between the p--Pb and pp samples is expected to introduce a 1\% effect on the \RpPb, which is negligible with respect to the other uncertainties.   
An interpolation procedure, which is described in Ref.~\cite{OurpPbPaper}, is adopted for the computation of the reference cross section in the last \pt interval, $10 < \pt < 14$~GeV/$c$, that was not measured in pp collisions at this energy. 
The interpolated value of $\textup{d}^2\sigma_{\rm pp}/\textup{d}y\textup{d}p_{\rm T}$ for this interval amounts to 
$10 \pm 2\ {\rm nb}/({\rm GeV}/c)$, 
where the quoted uncertainty refers to the total systematic uncertainty arising from the interpolation procedure and is uncorrelated with the uncertainties of the measured $\textup{d}^2\sigma_{\rm pp}/\textup{d}y\textup{d}p_{\rm T}$ for $\pt < 10$ GeV/$c$.

The estimated systematic uncertainties affecting the inclusive \PJpsi measurements are listed in Table~\ref{tab:systIncl}.
The dominant sources of uncertainty are related to the tracking and electron identification procedures. The remaining contributions are related to the signal extraction procedure, the \PJpsi input kinematic distributions used in the MC simulation, the dielectron decay channel branching ratio, and the integrated luminosity determination. 

\begin{table}[t]
\centering
	\caption{Summary of the systematic uncertainties, in percentage, of the inclusive \PJpsi cross section 
	${\rm d}^2\sigma_{\PJpsi} / {\rm d}y{\rm d}p_{\rm T}$ and nuclear modification factor \RpPb in different \pt intervals. 
	All contributions to the ${\rm d}^2\sigma_{\PJpsi} / {\rm d}y{\rm d}p_{\rm T}$ uncertainty are considered to be highly 
	correlated  over the \pt bins, 
	except that for the background subtraction which is considered as fully uncorrelated. 
The reported values for the measured  $\sigma_{\rm pp}$ reference cross section, which was determined up to $\pt = 10$~GeV/$c$~\cite{OurppRefPaper}, are the total uncertainties, both of statistical and systematic origins. The uncertainty of $\sigma_{\rm pp}$ for the interval 
$10 < \pt < 14$~GeV/$c$ 
is also the total uncertainty from the interpolation procedure as discussed in the text. 
} 
\begin{tabular}{c | r r r r r r r r }
\hline                    
  &\multicolumn{8}{c}{\pt \ (GeV/$c$)} \\
 \raisebox{1.3ex}{Source} & 0--1 & 1--2  & 2--3 & 3--4 & 4--5 & 5--7 & 7--10 & 10--14 \\ [0.5ex] 
\hline          
Tracking                 & 4.4 & 4.4 & 2.9 & 2.6 & 2.6 & 2.2 & 2.2 & 2.7  \\
PID                      & 1.0 & 1.0 & 1.2 & 1.3 & 1.1 & 2.1 & 4.4 & 6.1  \\
Signal shape             & 1.9 & 1.9 & 1.9 & 2.1 & 2.1 & 2.4 & 2.4 & 2.4  \\
Background subtraction   & 1.9 & 1.8 & 2.0 & 0.9 & 0.9 & 0.9 & 0.9 & 0.9  \\
MC input                 & 0.1 & 0.3 & 0.2 & 0.2 & 0.2 & 0.2 & 0.7 & 1.7  \\ 
$\sigma_{\rm pp}$        & 11.2 & 9.6 & 10.5 & 11.3 & 12.4 & 12.8 & 18.5 & 22.2  \\ 
Luminosity               & \multicolumn{8}{c}{3.7} \\
Branching ratio          & \multicolumn{8}{c}{0.5} \\ [0.5ex] 
\hline                     
Total (${\rm d}^2\sigma_{\PJpsi} / {\rm d}y{\rm d}p_{\rm T}$)         & 6.4 & 6.4 & 5.6 & 5.3 & 5.2 & 5.5 & 6.7 & 8.1  \\ [0.5ex]
Total (\RpPb)            & 12.9 & 11.5 & 11.9 & 12.5 & 13.4 & 13.9 & 19.7 & 23.7  \\ [1ex] 
\end{tabular}
\label{tab:systIncl}
\end{table}
The uncertainty of the tracking procedure dominates at low \pt values and is related to both the ITS-TPC matching efficiency and to the adopted track quality requirements.
The first component 
is estimated by evaluating the discrepancy in the matching probability of \TPC tracks to \ITS hits between data and Monte Carlo~\cite{Acharya:2017jgo}.
The observed discrepancy is used to re-scale the tracking efficiency of electrons in MC simulations in order to evaluate the difference in the resulting number of reconstructed \PJpsi candidates.
The second component is assessed by employing several variations to the adopted track selection criteria and by computing the RMS of the corrected \PJpsi yield distribution resulting after these variations.
The sum in quadrature of the uncertainties related to both these components is taken as systematic uncertainty on the tracking procedure.
The uncertainty related to the electron identification is estimated by evaluating the \TPC electron PID response for a clean sample of topologically identified electrons from conversion processes in data and computing the difference with the corresponding quantity from MC simulations. 
This per-track uncertainty is then propagated to the reconstructed \PJpsi candidates with the use of MC simulations.
The resulting uncertainty on the \PJpsi cross section
increases up to $\sim6\%$ towards high \pt values, where it is the largest uncertainty contribution. 
The systematic uncertainty related to the signal extraction procedure is due to both the background subtraction and the assumptions on the signal shape.
It is estimated as the RMS of the yield distributions corresponding to variations of the mass interval used for the signal counting, the alternative parameterisations employed to fit the correlated background, and the alternative method using the CB function to fit the signal.
The uncertainty on the signal shape ranges between $\sim$2\% at low \pt and $\sim$2.5\% at high \pt, 
whereas the uncertainty due to the background subtraction varies between~$\sim$1\% and~$\sim$2\% and is largest for the lowest \pt intervals.
The systematic uncertainty on the \PJpsi \pt distribution used as input for the computation of the efficiency corrections is determined by randomly varying, within one standard deviation contour, the parameters of a function fitted to the measured \pt distribution in p$-$Pb collisions~\cite{Adam:2015iga}, taking into account their correlations. The functional form used for this fit is discussed in Ref.~\cite{Bossu:2011qe} and very well describes the measured \PJpsi \pt distribution.  
The uncertainty of the integrated luminosity amounts to 3.7\% and is determined from the visible p$-$Pb cross sections measured in van der Meer scans as detailed in Ref.~\cite{Abelev:2014epa}.
Both the uncertainty on the integrated luminosity and that on the branching ratio ${\rm BR}\left( \PJpsi \rightarrow \ee \right) = (5.97 \pm 0.03)\%$~\cite{PDG2018} constitute global uncertainties for the inclusive \PJpsi cross section, fully correlated between all \pt intervals. All the other discussed sources of uncertainty are considered to be highly correlated~\footnote{With high correlation, we  mean a Pearson coefficient larger than 0.7.} over \pt, with the exception of the background uncertainty, which is considered as uncorrelated.

The total relative uncertainty for the reference pp cross section, $\sigma_{\rm pp}$, is also reported in 
Table~\ref{tab:systIncl}. In this case, the values up to $\pt=10$~GeV/$c$, are the total uncertainties, of both  statistical and systematic origin,  associated to the measurement performed up to $\pt=10$~GeV/$c$~\cite{OurppRefPaper}, while that  for the $10 < \pt < 14$~GeV/$c$ interval is the relative uncertainty of the interpolated cross section 
($10 \pm 2\ {\rm nb}/({\rm GeV}/c)$) quoted before. The uncertainty of $\sigma_{\rm pp}$ propagates only to the \RpPb\ observable  
and it 
varies between $\sim$10\% and $\sim$20\% and is largest for the highest \pt intervals.

\subsection{Determination of the non-prompt \PJpsi fraction}

The fraction \fb of the \PJpsi yield originating from b-hadron decays is measured for $\pt > 1$~GeV/$c$ by discriminating, on a statistical basis, the reconstructed \PJpsi candidates according to the displacement between their production vertex and the primary p$-$Pb collision vertex.
The discrimination is realized by means of an unbinned two-dimensional likelihood fit, following the same technique adopted in previous analyses for the pp~\cite{Abelev:2012gx}, ~p$-$Pb~\cite{OurpPbPaper}, and \PbPb~\cite{Adam:2015rba} systems. In particular, it is performed by maximising the following log-likelihood function 

\begin{equation}
    {\rm ln} L 
    =\sum_{1}^{N}{\rm ln}\left[f_{\rm Sig}\times F_{\rm Sig}(x)\times M_{\rm Sig}(m_{\rm e^+e^-}) + (1-f_{\rm Sig})\times F_{\rm Bkg}(x)\times M_{\rm Bkg}(m_{\rm e^+e^-})\right] \ , 
    \label{EqLikelihood}
\end{equation}

in which $N$ indicates the number of \ee\ pairs within the $2.32 < m_{\rm e^+e^-} < 4.00$ GeV/$c^2$ invariant mass interval. The pseudoproper decay length $x$ is introduced to separate 
\PJpsi originating from the decay of b-hadrons
from prompt \PJpsi. It is defined as
\begin{equation}
    x = \frac{c\times L_{\rm xy} \times m_{\PJpsi}}{\pt}\ ,
\end{equation}
where $c$\ is the speed of light,  $L_{\rm xy} = \Vec{L}\cdot \Vec{p_{\rm T}}/\pt$ is the signed projection of the pair flight distance, $\Vec{L}$, onto its transverse momentum vector, $\Vec{\pt}$, and $m_{\PJpsi}$ is the \PJpsi pole mass value~\cite{PDG2018}.
The terms $F_{\rm Sig}$ ($F_{\rm Bkg}$) and $M_{\rm Sig}$ ($M_{\rm Bkg}$) in Eq.~\ref{EqLikelihood} represent the probability density functions (PrDFs) describing the signal (background) pair distributions as a function of $x$ and $m_{\rm e^+e^-}$, respectively, whereas $f_{\rm Sig}$ denotes the ratio of signal to all candidates within the considered mass interval. The signal $x$ PrDF is given by  
\begin{equation}
   F_{\rm Sig}(x) = f'_{\rm b}\times F_{\rm b}(x) + (1-f'_{\rm b})\times F_{\rm prompt}(x),
\end{equation}
with $F_{\rm prompt}(x)$ and $F_{\rm b}(x)$ indicating the prompt and non-prompt \PJpsi PrDFs, and $f'_{\rm b}$ being the uncorrected fraction of \PJpsi coming from b-hadron decays.

\begin{figure}[t]
    \begin{center}
    \includegraphics[width = 0.497\textwidth]{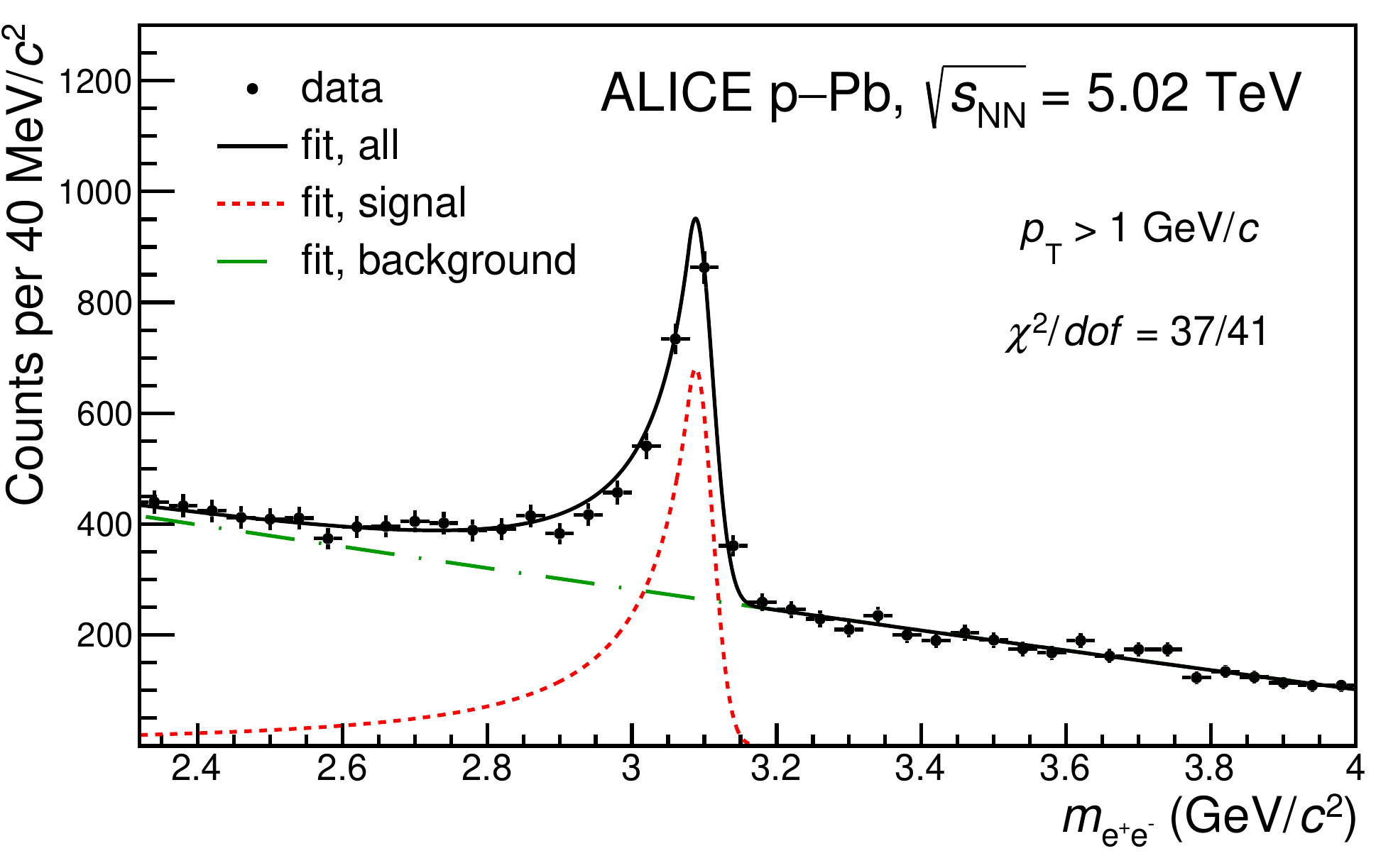}
    \includegraphics[width = 0.497\textwidth]{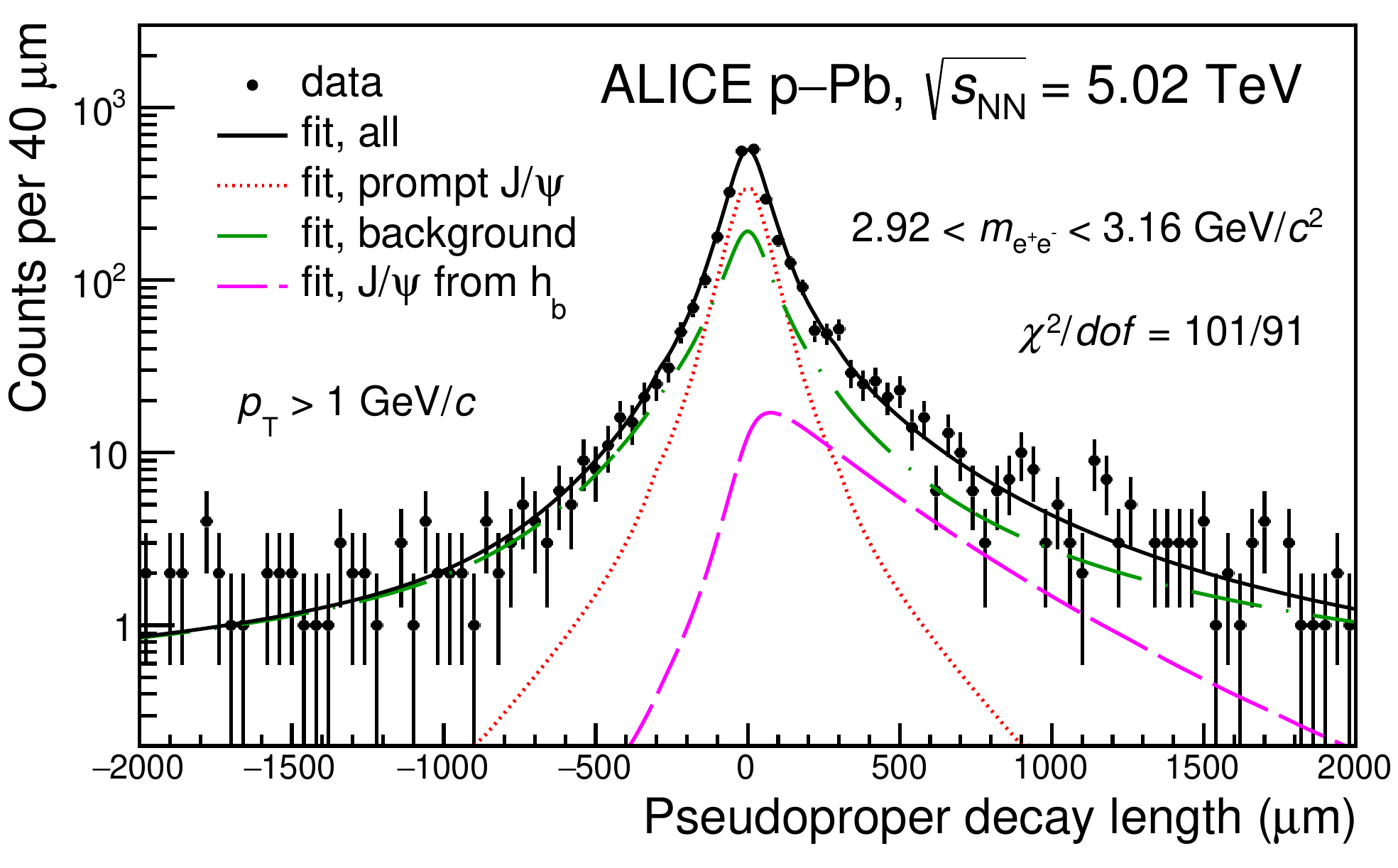}
    \end{center}
    \caption{Invariant mass (left panel) and pseudoproper decay length (right panel) distributions for \PJpsi candidates with $\pt > 1$ GeV/$c$. The latter distribution is limited to the \PJpsi candidates under the signal peak region $2.92~<~m_{\rm e^+e^-}~<~3.16$~GeV/$c^2$. 
    The projections of the maximum likelihood fit functions are shown superimposed
    and the $\chi^2$ values of these projections are also reported for both distributions.
    }
    \label{fig:TechnicalNonprompt}
\end{figure}

The evaluation of the different PrDFs used in Eq.~\ref{EqLikelihood} is performed relying either on data or on MC simulations 
and following the same procedures described in previous analyses~\cite{Abelev:2012gx,Adam:2015rba}.
For the MC simulations, the prompt \PJpsi component was generated  
with the  \pt and $y$ distributions obtained with a procedure analogue to that previously discussed for the inclusive \PJpsi analysis, while the non-prompt \PJpsi component was obtained using PYTHIA 6.4~\cite{PYTHIA} with Perugia-0 tuning~\cite{PERUGIA0} to simulate the production of beauty hadrons.
Also in this case, the \PJpsi decay into dielectrons was simulated using the EvtGen package~\cite{EVTGEN} in combination with the PHOTOS model~\cite{PHOTOS} in order to provide a proper description of the radiative decay channel. 
The background $x$ PrDF, 
$F_{\rm Bkg}(x)$, is determined in three invariant mass ranges 
by fitting the $x$ distributions of dielectron candidates in the lower ($2.32 < m_{\rm e^+e^-} < 2.68$ GeV/$c^2$) and upper ($3.20 < m_{\rm e^+e^-} < 4.00$ GeV/$c^2$) side bands of the invariant mass distributions and by interpolating the resulting fit functions to the region under the invariant mass signal peak ($2.68 < m_{\rm e^+e^-} < 3.20$ GeV/$c^2$). 
The experimental resolution function, $R(x)$, which is the key ingredient in the $F_{\rm prompt}(x)$, $F_{\rm b}(x)$ and $F_{\rm Bkg}(x)$ PrDFs, is 
evaluated from the $x$ distributions of prompt \PJpsi in MC simulations, reconstructed after applying the same selection criteria as in data.  
In order to improve the resolution of the secondary decay vertices, it is required that at least one of the two \PJpsi candidate decay tracks has a hit in the innermost SPD layer.
A tune-on-data procedure~\cite{OurpPbPaper} is applied to the MC sample in order to reproduce the observed single-track impact parameter distributions. This minimises the discrepancy between data and simulation, reducing the systematic uncertainty related to the $R(x)$ determination.
The $F_{\rm b}(x)$ PrDF is obtained as the convolution of the $R(x)$ 
function and a template of the $x$ distribution for the mixture of b-hadrons decaying into \PJpsi. The latter is obtained with a MC simulation study of the kinematics of the b-hadron decays. In this simulation, the \pt-distribution of the b-hadrons is obtained from pQCD calculations at fixed order with next-to leading-log re-summation (FONLL)~\cite{FONLL}. The decay description is based on the EvtGen package~\cite{EVTGEN}, and the relative abundance of b-hadron species as a function of \pt is based on the precise measurements reported by the LHCb collaboration in pp collisions~\cite{Aaij:2019pqz}, which are consistent with those measured in p--Pb collisions~\cite{Aaij:2019lkm}.  
The $x$ resolution estimated from the MC simulations is characterised by a pronounced dependence as a function of the \PJpsi \pt: 
for events with both \PJpsi decay tracks yielding a hit in the first SPD layer, the RMS of the $R(x)$ distribution 
ranges from $\sim$140 $\mu$m at $p_{\rm T} = 1.5$ GeV/$c$ to $\sim$50 $\mu$m at $p_{\rm T} > 7$ GeV/$c$.
This allows the fraction of non-prompt \PJpsi to be determined for events with \PJpsi \pt greater than 
1~GeV/$c$ as well as in five transverse momentum intervals (1--3, 3--5, 5--7, 7--10 and 10--14 GeV/$c$).
The projections of the maximised likelihood fit function superimposed over the $m_{\rm e^+e^-}$ and $x$ distributions of \PJpsi candidates with $\pt > 1$ GeV/$c$ are shown as an example in Fig.~\ref{fig:TechnicalNonprompt}. 

The \fb fraction is obtained after correcting the $f'_{\rm b}$ values (Eq.~\ref{EqCorr} ) to account for slightly different $\left\langle A\times \epsilon \right\rangle$ factors of prompt and non-prompt \PJpsi:

\begin{equation}
    f_{\rm b} = \left( 1 + \frac{1-f'_{\rm b}}{f'_{\rm b}} \times \frac{\left\langle A\times \epsilon \right\rangle_{\rm B}}{\left\langle A\times \epsilon \right\rangle_{\rm prompt}}  \right)^{-1}\ .
     \label{EqCorr}
\end{equation}

This small correction is computed relying on MC simulations assuming prompt \PJpsi to be unpolarised.
A small residual polarisation, resulting from the admixture of the different b-hadron decay channels, is assumed for non-prompt \PJpsi as predicted by EvtGen~\cite{EVTGEN}.
Under these conditions, the correction mainly  originates from the difference in the \pt distribution between the two components and
is found to be significant only for the \pt-integrated case.
Small relative variations of the corrected \fb values, in the order of $\sim$1--4\%, are expected in the case of a null-polarisation assumption for non-prompt \PJpsi~\cite{OurpPbPaper}. The variations estimated in extreme polarisation scenarios for prompt \PJpsi are discussed in Ref.~\cite{Abelev:2012gx}.
Considering the null or very small degree of polarisation measured in pp collisions at the LHC~\cite{Aaij:2011jh,Abelev:2011md,Chatrchyan:2013cla,Acharya:2018uww}, these variations are not further propagated to the final results, and only the choice of the \pt shapes used as input for the MC simulations is taken into account for the systematic uncertainty evaluation, as discussed below.

The evaluated systematic uncertainties affecting the measurements of \fb in the five \pt intervals as well as in the \pt-integrated range ($\pt > 1$ GeV/$c$) are listed in Table~\ref{tab:hresultNonPrompt}. 
Most of the listed contributions are due to incomplete knowledge of the different PrDFs used as input for the likelihood fits. 
An additional contribution originates from the assumptions on the 
\pt distributions 
employed for the computation of the correction factor of Eq.~\ref{EqCorr}. 
The uncertainties affecting the evaluation of the resolution function and of the background PrDF  constitute the largest contributions to the total systematic uncertainty of the \fb measurements.
The former is estimated by propagating to the $R(x)$ PrDF the residual discrepancy of the single-track impact parameter distributions between data and MC simulations after the application of the previously discussed tuning procedure.
The latter is evaluated by repeating the likelihood fits after varying the procedure used for the determination of the $F_{\rm Bkg}(x)$ PrDFs, following the same approach described in Ref.~\cite{Adam:2015rba}.
Both uncertainties increase towards low transverse momenta and are largest for the lowest \pt interval, where they amount to 10\% and 8.5\%, respectively.
The uncertainty related to the invariant mass PrDF of the \PJpsi signal is estimated by changing the width of the CB function used to parameterise the $M_{\rm Sig}$ PrDF so as to vary the fraction of signal enclosed within the $2.92 < m_{\rm e^+e^-} < 3.16$~GeV/$c^2$ interval by $\pm$2.5\%. 
The likelihood fits are repeated, and the variation of the resulting \fb values
is taken as systematic uncertainty. 
The uncertainty related to the invariant mass background PrDF is estimated as the RMS of the \fb value distributions obtained after employing different parameterisations and alternative fitting approaches for the evaluation of the $M_{\rm Bkg}$ PrDF. 
The estimate of the systematic uncertainty affecting the non-prompt \PJpsi $x$ PrDF is performed by repeating the likelihood fits after employing PYTHIA 6.4 for the description of the \pt-distribution, decay kinematics, and relative abundance of the beauty hadrons in the MC simulations used to model the $F_{\rm b}(x)$ PrDF.
The relative variation of the resulting \fb values, assumed as systematic uncertainty, increases up to $\sim$3\% towards low transverse momenta. This uncertainty also contemplates any conservative assumption for a rapidity dependence of the relative abundances of beauty hadrons at the LHC.
The uncertainty related to the acceptance times efficiency correction procedure is assessed by testing different hypothesis for the kinematic \pt-spectra used to compute the $\left\langle A\times \epsilon \right\rangle$ values that enter into Eq.~\ref{EqCorr}.
Among the tested variations, a tune-on-data parameterisation based on the  Run 1 measurement~\cite{OurpPbPaper} for prompt \PJpsi, a \pt-distribution based on FONLL calculations~\cite{FONLL} for the non-prompt component, and the inclusion or exclusion of nuclear shadowing modifications according to the EPPS16 parameterisation~\cite{EPPS16} are considered.
The resulting variations of the correction factors are largest for the \pt-integrated measurement, where they amount to $\sim$3\%, while they are smaller than $\sim$1\% within the analysed \pt-intervals.
The overall systematic uncertainty of the $f_{\rm b}$ measurements is found to increase up to 13.7\% towards low transverse momenta, mostly as a consequence of both the increasing combinatorial background and the worsening of the $x$ resolution.

\begin{table}[t]
\centering
\caption{List of the systematic uncertainties (in percent) for  
the fraction of \PJpsi from b-hadron decays in the different analysed \pt intervals. The symbol ``$-$" denotes a negligible contribution.} 
\begin{tabular}{ c | c c c c c c }
  &\multicolumn{6}{c}{\raisebox{0.5ex}{\pt (GeV/$c$)}} \\
  \cline{2-7}
\raisebox{1.3ex}{Source} & $ > 1$ & 1--3 & 3--5 & 5--7 & 7--10 & 10--14 \\ [0.5ex] 
\hline          
Resolution function                & 6.3 & 10.0 & 5.5 & 2.4 & 1.3 & 1.1   \\
$x$ PrDF of non-prompt \PJpsi      & 1.7 & 3.2  & 0.9 & 0.5 & 0.3 & 0.3   \\
$x$ PrDF of background             & 5.0 & 8.5  & 3.2 & 3.2 & 2.6 & 1.8   \\
Invariant mass PrDF of signal      & 1.4 & 1.1  & 1.3 & 1.5 & 1.8 & 1.8   \\ 
Invariant mass PrDF of background  & 1.5 & 1.8  & 1.4 & 1.2 & 0.8 & 1.2   \\ 
Acceptance $\times$ efficiency     & 3.0 & 0.3  & 0.7 & 0.2 &  $-$  & $-$    \\  [1ex]
\hline                     
Total                              & 9.0 & 13.7 & 6.7 & 4.5 & 3.5 & 3.0   \\ [0.5ex]
\end{tabular}
\label{tab:hresultNonPrompt}
\end{table}

The prompt and non-prompt \PJpsi nuclear modification factors are computed by combining the measurements of \fb with the previously discussed nuclear modification factors $R_{\rm pPb}^{\rm incl.\ \PJpsi}$ of inclusive \PJpsi:

\begin{equation}
    R_{\rm pPb}^{\text{non-prompt}\ \PJpsi} = \frac{f_{\rm b}^{\rm pPb}}{f_{\rm b}^{\rm pp}}\ R_{\rm pPb}^{\rm incl.\ \PJpsi}\ ,\ \ \ \ \ \ \ \ \ \ \ \ \ \ \ 
    R_{\rm pPb}^{\text{prompt}\ \PJpsi} = \frac{1-f_{\rm b}^{\rm pPb}}{1-f_{\rm b}^{\rm pp}}\ R_{\rm pPb}^{\rm incl.\ \PJpsi}\ .    
    \label{eqRpPbfB}
\end{equation}

The value of \fb in pp collision at $\sqrt{s} = 5.02$ TeV, indicated as $f_{\rm b}^{\rm pp}$ in Eq.~\ref{eqRpPbfB}, is determined by means of the same interpolation procedure adopted in previous analyses for the~p$-$Pb~\cite{OurpPbPaper} and \PbPb~\cite{Adam:2015rba} systems.
The procedure consists of fitting to existing midrapidity \fb measurements at $\sqrt{s} = 1.96$ TeV (from CDF~\cite{CDF1}) and $\sqrt{s} = 7$ TeV (from  ALICE~\cite{Abelev:2012gx}, ATLAS~\cite{Aad:2011sp}, and CMS~\cite{Khachatryan:2010yr}) the semi-phenomenological function discussed in Ref.~\cite{Adam:2015rba}, which includes FONLL predictions~\cite{FONLL} for the non-prompt \PJpsi production cross section.
An energy interpolation is then performed to derive the $f_{\rm b}^{\rm pp}(\pt)$ at $\sqrt{s} = 5.02$ TeV as a function of \pt.
The average value of $f_{\rm b}^{\rm pp}$ in a given \pt interval is obtained by weighting $f_{\rm b}^{\rm pp}(\pt)$ over the inclusive \PJpsi spectrum in pp collisions in that \pt interval.  
Compared to our previous estimates~\cite{OurpPbPaper}, a tune-on-data spectrum based on the inclusive \PJpsi yield measured by ALICE in pp collisions at $\sqrt{s} = 5.02$ TeV~\cite{OurppRefPaper} is now employed for this purpose.
The values of $f_{\rm b}^{\rm pp}$ at $\sqrt{s} = 5.02$ TeV, computed in the considered momentum intervals, are reported in Table~\ref{tab:fbpp}.
The quoted uncertainties take into account the uncertainties of both data and FONLL predictions, as well as an additional systematic uncertainty due to the choice of the functional form (either a linear, or an exponential, or a power law function) employed for the energy interpolation procedure.

\begin{table}[t]
\centering
\caption{Fraction of non-prompt \PJpsi in pp collisions at $\sqrt{s} = 5.02$ TeV computed in different transverse momentum intervals. The reported values and uncertainties are derived following the interpolation procedure detailed in the text and in Ref.~\cite{Adam:2015rba}.} 
\begin{tabular}{ c | c }
\pt (GeV/$c$)                & $f_{\rm b}^{\rm pp}$ at $\sqrt{s} = 5.02$ TeV   \\ [0.5ex]
\hline          
$> 0$    & $0.135 \pm 0.013 $  \\
1--3             & $0.117 \pm 0.013$   \\
3--5             & $0.144 \pm 0.012$   \\
5--7             & $0.188 \pm 0.014$   \\
7--10            & $0.246 \pm 0.019$   \\  
10--14           & $0.333 \pm 0.038$   \\  [0.5ex]
\end{tabular}
\label{tab:fbpp}
\end{table}


\section{Results}

\begin{figure}[h]
    \begin{center}
    \includegraphics[width = 0.6\textwidth]{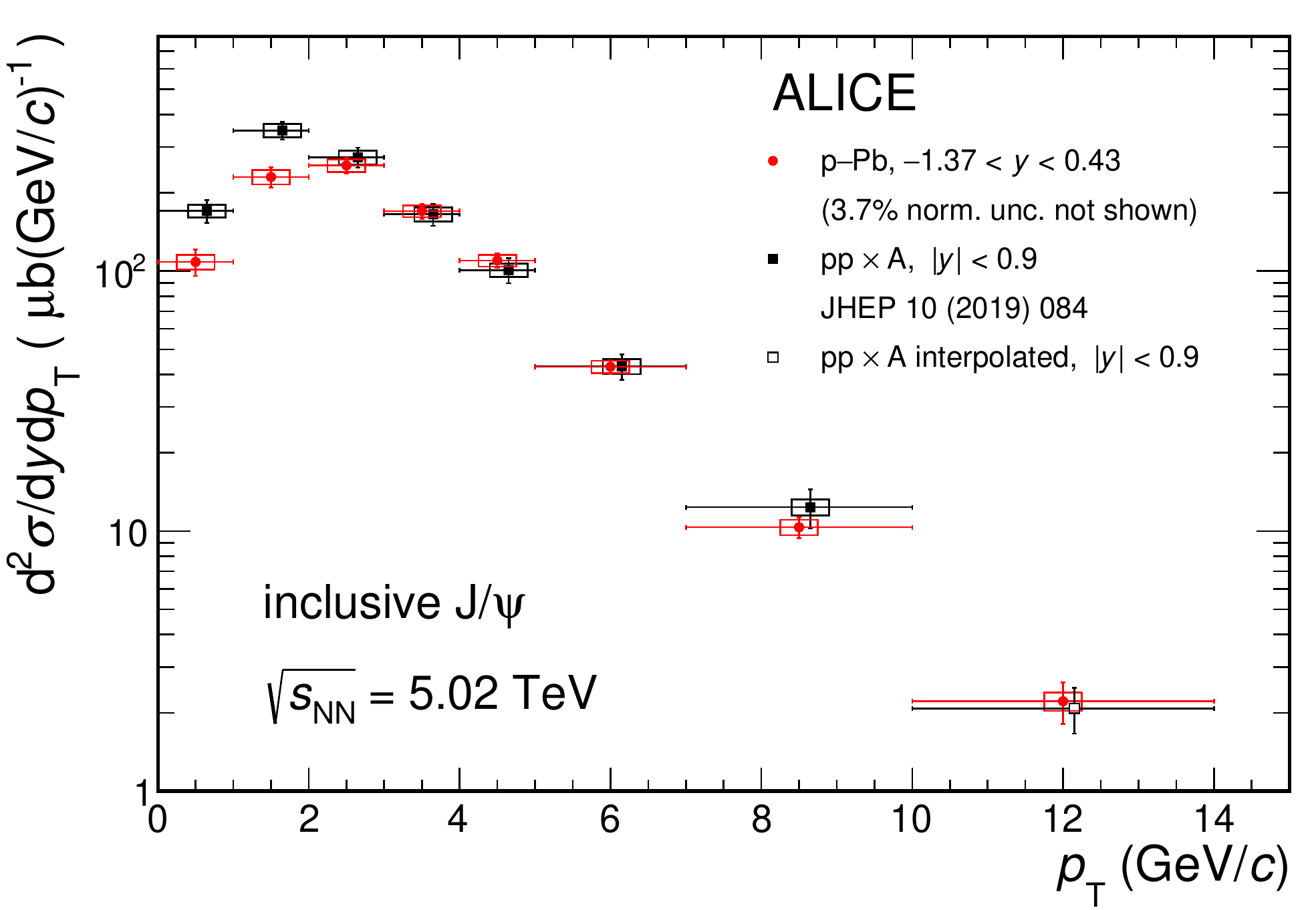}
    \end{center}
    \caption{The \pt-differential inclusive \PJpsi\ cross section per unit of rapidity, 
    $\textup{d}^2\sigma_{\rm incl.\,\PJpsi}/\textup{d}y\textup{d}p_{\rm T}$, as a function of \pt in p--Pb collisions (red circles) compared 
    with the analogous cross section measured in pp collisions at $\sqrt{s}=5.02$~TeV~\cite{OurppRefPaper} multiplied by  
    the Pb mass number (${\rm A}=208$) (black closed squares, shifted horizontally by 150 MeV/$c$ for better visibility). 
    The vertical error bar and the box on top of each point represent the statistical and systematic uncertainty, respectively. 
    The open square symbol (also shifted by 150 MeV/$c$) shows the value for pp collisions in the \pt interval 10-14~GeV/$c$, which was obtained with the interpolation procedure. In this case the error bar corresponds to the total uncertainty.}
    \label{fig:Inclusive}
\end{figure}

The inclusive \PJpsi cross section is measured in $-1.37<y<0.43$ both for $\pt >0$ and differentially in \pt considering seven \pt intervals, with the first and last bins 
being 
[0$-$1]~GeV/$c$ and [10$-$14]~\GeV/$c$, respectively.
The value of the \pt-integrated inclusive \PJpsi cross section per unit of rapidity is   $\textup{d}\sigma/\textup{d}y = \rm 999 \pm 33\ (stat.) \pm 56\ (syst.)\ \mu$b.
The \pt-differential cross section of inclusive \PJpsi per unit of rapidity,
$\textup{d}^2\sigma_{\rm incl.\,\PJpsi}/\textup{d}y\textup{d}p_{\rm T}$, 
is shown in Fig.~\ref{fig:Inclusive} in comparison with the cross section measured in pp collisions at $\sqrt{s}=5.02$~TeV~\cite{OurppRefPaper} multiplied by ${\rm A}=208$. The latter extends up to $\pt=10$~GeV/$c$. The highest \pt point for pp collisions, which is shown in the figure with the empty symbol, was obtained using the interpolation procedure as described before. 

\begin{figure}[h]
    \begin{center}
    \includegraphics[width = 0.6\textwidth]{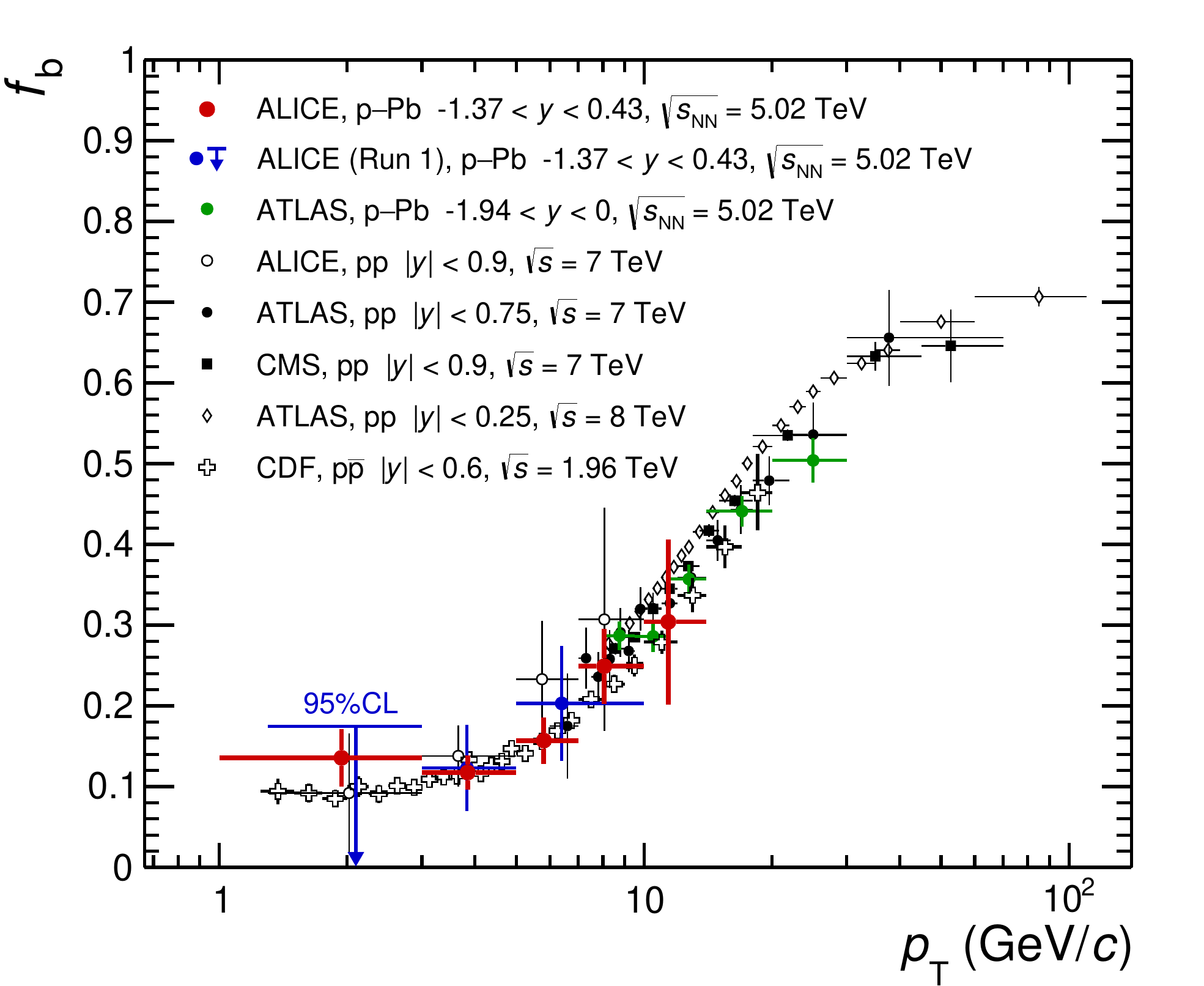}
    \end{center}
    \caption{Fraction of \PJpsi from b-hadron decays at midrapidity as a function of the \PJpsi \pt in p--Pb collisions at $\snn = 5.02$ TeV (red closed circles) compared with results from the ALICE~\cite{OurpPbPaper} and ATLAS~\cite{Aad:2015ddl} collaborations in the same collision system (blue and green closed circles, and the blue arrow that shows an upper limit at 95\%~CL in the range $1.3<\pt<3$~GeV/$c$). 
    The results from CDF~\cite{CDF1} in p$\rm \overline{p}$ collisions at $\sqrt{s} = 1.96$ TeV and those of the ALICE~\cite{Abelev:2012gx}, ATLAS~\cite{Aad:2011sp,Aad:2015duc}, and CMS~\cite{Khachatryan:2010yr} collaborations in pp collisions at $\sqrt{s} = 7$ TeV or $\sqrt{s} = 8$ TeV are also shown (black symbols).
    For all experiments, vertical error bars represent the quadratic sum of the statistical and systematic uncertainties.
    The data points of ALICE are placed horizontally at the mean value of the \pt distribution within each \pt-interval, determined from the MC simulations described in the text.}
    \label{fig:fb}
\end{figure}

The fraction of \PJpsi from b-hadron decays in the kinematic range 
$\pt > 1$ GeV/$c$ and $-1.37 < y < 0.43$, which is referred to as "visible region" in the following, is found to be
$f_{\rm b} = \rm 0.125 \pm 0.017\;{\rm (stat.)}  \pm 0.011\;{\rm (syst.)  }$, 
where the first quoted uncertainty is statistical
and the second one is systematic. 
The \fb measurements in the five analysed \pt intervals are shown in Fig.~\ref{fig:fb} in comparison with our previous results~\cite{OurpPbPaper} and with the results from the ATLAS collaboration~\cite{Aad:2015ddl}, measured for $\pt > 8$ GeV/$c$ within a similar rapidity interval ($-1.94 < y < 0$).
The measurements from the CDF~\cite{CDF1}, ATLAS~\cite{Aad:2015duc}, and CMS~\cite{Khachatryan:2010yr} experiments in pp and p$\rm \overline{p}$ collisions at midrapidity are also shown for comparison.  
With respect to our previous results~\cite{OurpPbPaper}, the present measurements are performed over a wider \pt range, with a more granular binning, and show a significantly improved precision, with about half of the statistical uncertainty within similar \pt intervals.

The prompt and non-prompt \PJpsi production cross sections are obtained from the combination of the \fb fractions with the measurements 
of the inclusive \PJpsi cross section $\sigma_{\rm \PJpsi}$:
\begin{linenomath}\begin{equation}
    \sigma_{\rm \PJpsi\ from\ h_{\rm b}} = f_{\rm b}\times\sigma_{\rm \PJpsi},\ \ \ \ \ \ \ \ \ 
    \sigma_{\rm prompt\ \PJpsi} = (1-f_{\rm b})\times\sigma_{\rm \PJpsi}.
    \label{eq:XsecVis}
\end{equation}\end{linenomath}

The non-prompt \PJpsi cross section in the visible region, $\sigma_{\rm \PJpsi\ from\ h_{\rm b}}^{\rm vis} = \rm 201 \pm 28\ (stat.) \pm 21\ (syst.) \ {\rm \mu b}$, is computed using the inclusive \PJpsi cross section for $\pt >1$ GeV/$c$, which amounts to $\rm 1603 \pm 55\ (stat.) \pm 89\ (syst.) \ {\rm \mu b}$.

In order to derive the \pt-integrated values of the prompt and non-prompt \PJpsi cross section at midrapidity, $\sigma_{\rm \PJpsi\ from\ h_{\rm b}}^{\rm vis}$ is extrapolated down to $\pt = 0$ following the approach described in our previous work~\cite{OurpPbPaper}.
The extrapolation is performed assuming the shape of the \pt distribution of b-quarks obtained from FONLL~\cite{FONLL} with the CTEQ6.6 PDFs~\cite{Nadolsky:2008zw} modified according to EPPS16 nPDF parameterisation~\cite{EPPS16}. 
The fragmentation of b-quarks into hadrons is then modelled using PYTHIA 6.4~\cite{PYTHIA} with the Perugia-0 tune~\cite{PERUGIA0}. 
The ratio of the extrapolated cross section for 
$p_{\rm T} > 0$
and $-1.37 < y < 0.43$ to that in the visible region 
($p_{\rm T} > 1$~GeV/$c$ and $-1.37 < y < 0.43$) 
equals 
$1.127^{+0.014}_{-0.025}$
, where the quoted uncertainty takes into account the FONLL, CTEQ6.6 and EPPS16 uncertainties, as described in Ref.~\cite{OurpPbPaper}, as well as an additional uncertainty, which is related to that on the relative abundance of beauty hadron species in the extrapolated range. 
The latter is estimated to be about 0.4\% after changing the assumed fractions of beauty hadrons according to the recent LHCb measurements~\cite{Aaij:2019pqz}. Also, in this case, the considered variation largely includes a possible dependence of these fractions on rapidity.
Thus, the measured cross section corresponds to more than 85\% of the \pt-integrated cross section at midrapidity.
Dividing by the rapidity range $\Delta y = 1.8$, the following value is derived for the non-prompt \PJpsi cross section per unit of rapidity 
($\pt>0$ and $-1.37<y<0.43$):
\begin{linenomath}
\begin{equation*}
    \frac{\textup{d}\sigma_{\rm \PJpsi\ from\ h_{\rm b}}}{\textup{d}y} = 
    \rm 125.6 \pm 17.6\ (stat.) \pm 13.3\ (syst.)\ ^{+1.6}_{-2.8}\ (extr.)\ \mu b.
\end{equation*}
\end{linenomath}
The corresponding value for the prompt component is obtained as the difference between 
the inclusive \PJpsi cross section, which is measured for $\pt > 0$, and that of \PJpsi from b-hadron decays, as determined with the extrapolation procedure described above. It is ($\pt>0$ and $-1.37<y<0.43$)   
\begin{linenomath}
\begin{equation*}
    \frac{\textup{d}\sigma_{\rm prompt\ \PJpsi}}{\textup{d}y} = 
    \rm 873.1 \pm 33.6\ (stat.) \pm 50.4\ (syst.)\ ^{+1.6}_{-2.8}\ (extr.)\ \mu b.
\end{equation*}
\end{linenomath}

\begin{figure}[t]
    \begin{center}
    \includegraphics[width = 0.495\textwidth]{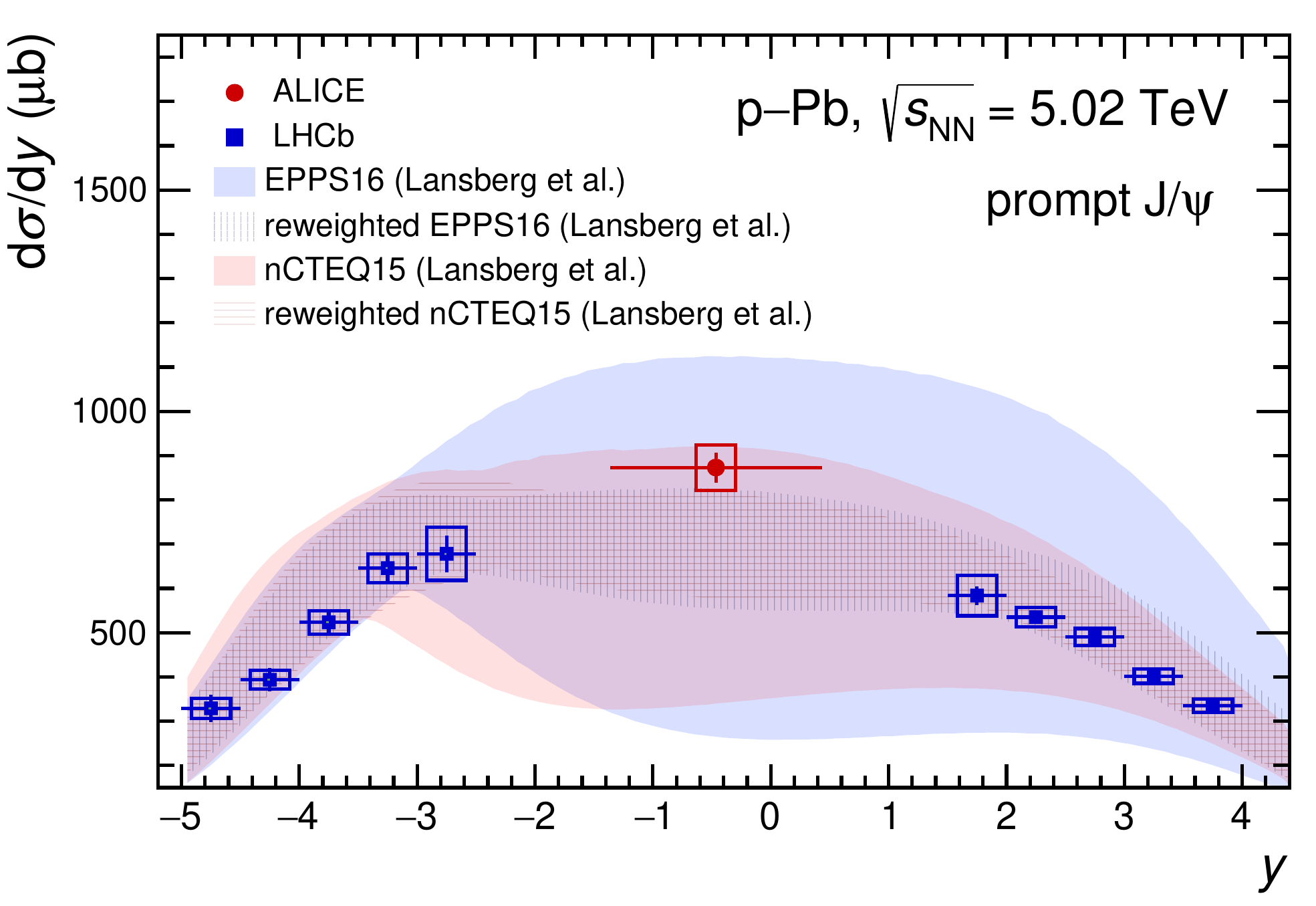}
     \includegraphics[width= 0.495\textwidth]{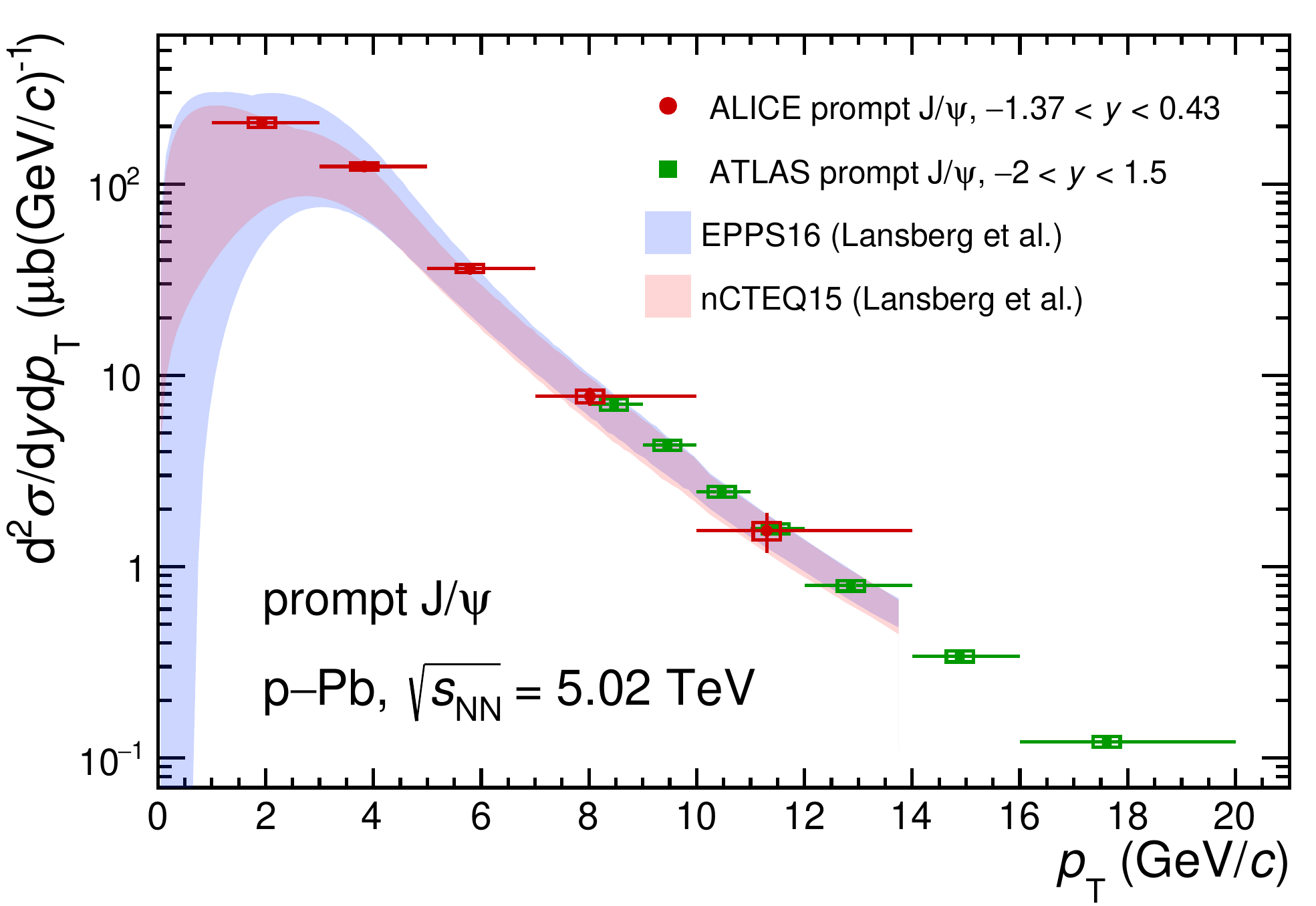}
    \end{center}
    \caption{$\textup{d}\sigma_{\rm prompt\ \PJpsi}/\textup{d}y$ as a function of the rapidity in the centre-of-mass frame (left panel) as obtained in this work at midrapidity and by the LHCb collaboration in the forward and backward rapidity regions~\cite{Aaij:2013zxa}, and $\textup{d}^2\sigma_{\rm prompt\ \PJpsi}/\textup{d}y\textup{d}p_{\rm T}$ as a function of \pt (right panel) compared 
    with ATLAS results~\cite{Aaboud:2017cif} (reported up to $\pt = 20$ GeV/$c$). 
    The vertical error bars and the boxes on top of each point represent the statistical and systematic uncertainties. 
    In the left panel, the systematic uncertainty of the ALICE data point includes also the contribution from the extrapolation procedure to go from the visible region ($\pt>1$~GeV/$c$) to $\pt>0$, as described in the text.
    For the measurements as a function of \pt, the data symbols 
    are placed within each bin at the mean of the \pt  distribution determined from MC simulations. 
    The results of a model~\cite{Lansberg:2016deg,Kusina:2017gkz,Shao:2012iz,Shao:2015vga} including nuclear shadowing based on the  
 EPPS16~\cite{EPPS16} and nCTEQ15~\cite{nCTEQ15} nPDFs are shown superimposed on both panels (see text for details). 
 In the right panel, the computations refer only to the ALICE rapidity range.}
    \label{fig:xsecprompt}
\end{figure}

\begin{figure}[t]
    \begin{center}
    \includegraphics[width = 0.496\textwidth]{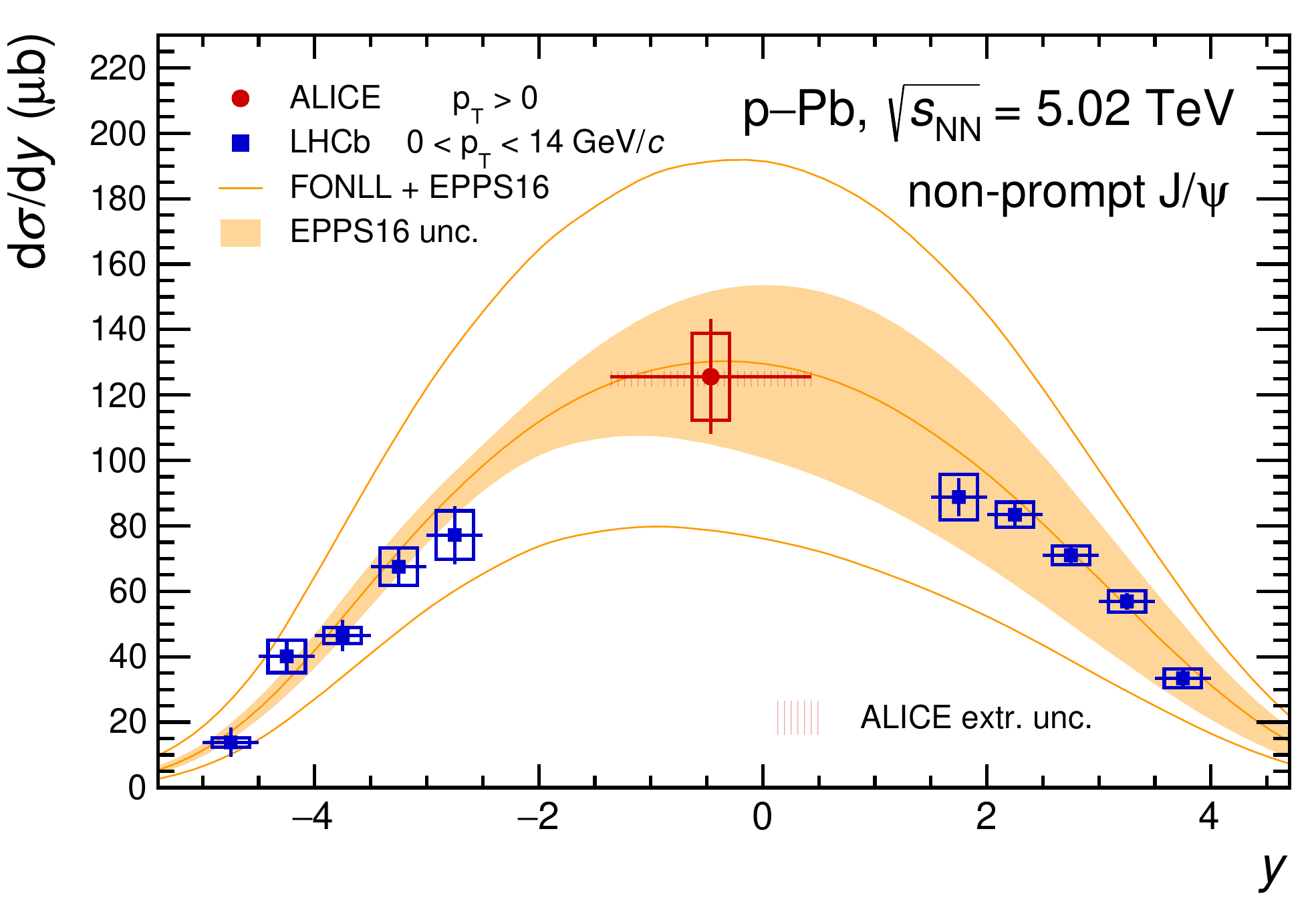}
    \includegraphics[width = 0.496\textwidth]{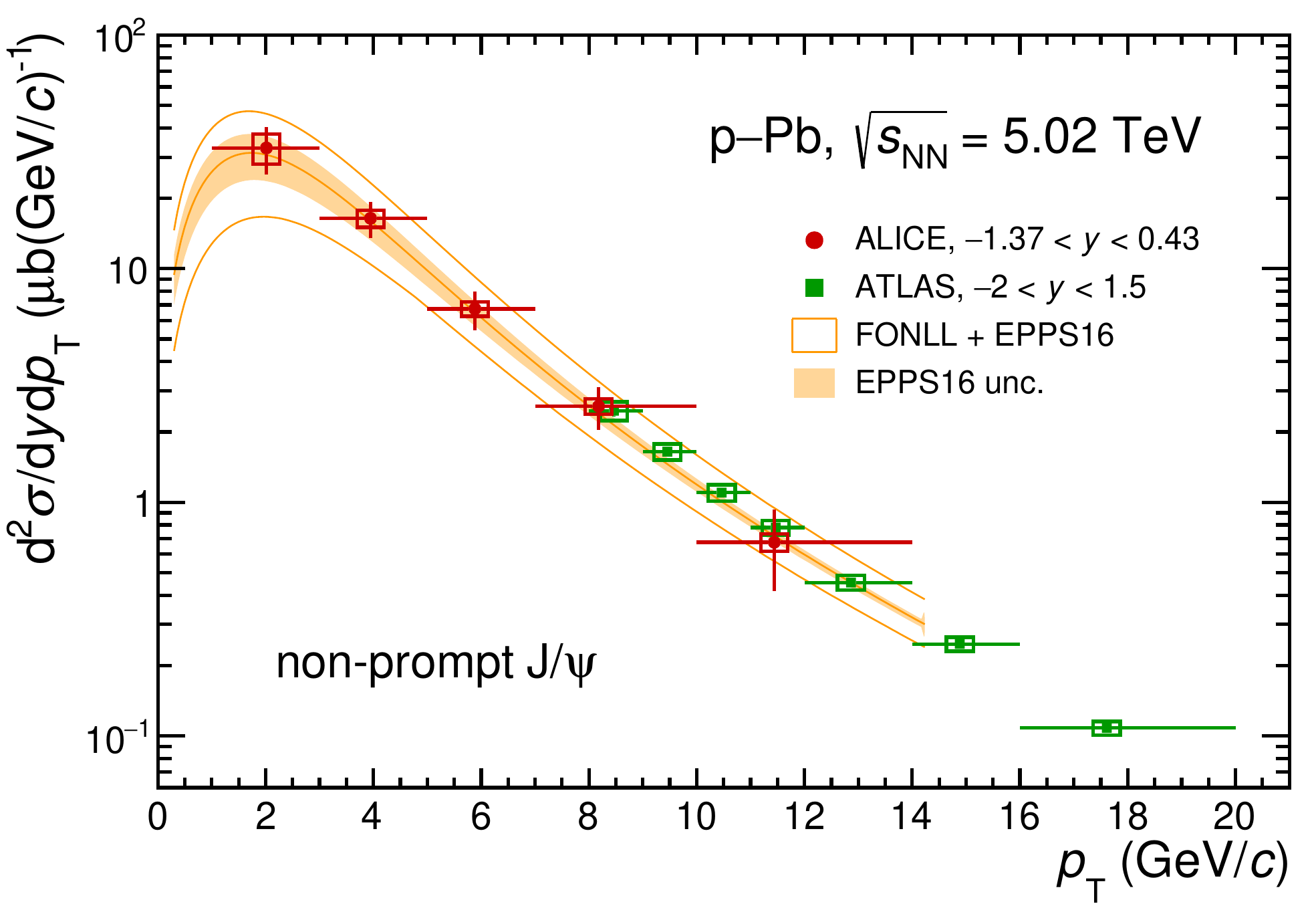} 
    \end{center}
    \caption{
    $\textup{d}\sigma_{\rm \PJpsi\ from\ h_{\rm b}}/\textup{d}y$ as a function of  
    rapidity 
    (left panel) as obtained at midrapidity in this work for $\pt>0$ and by the  LHCb collaboration in the forward and backward rapidity regions~\cite{Aaij:2013zxa} for $0<\pt<14$~GeV/$c$, and $\textup{d}^2\sigma_{\rm \PJpsi\ from\ h_{\rm b}}/\textup{d}y\textup{d}p_{\rm T}$ as a function of \pt (right panel) compared with ATLAS measurements~\cite{Aaboud:2017cif} (shown up to $\pt = 20$ GeV/$c$). 
    The statistical and systematic uncertainties are shown as vertical error bars and hollow boxes, respectively. The extrapolation uncertainty related to the procedure to go from the visible region 
	($\pt > 1$~GeV/$c$) to $\pt > 0$ for the ALICE data point in the left panel is indicated with a dashed band.
For the measurements as a function of \pt, data symbols are placed at the mean value of the \pt distribution within each bin.
    The results are compared to FONLL computations~\cite{FONLL} with EPPS16~\cite{EPPS16} nPDFs, highlighting the total theoretical uncertainty (empty band) and the contribution from EPPS16 (coloured band). In the right panel, model computations are obtained in the same rapidity range of the ALICE results, namely $-1.37<y<0.43$.}
    \label{fig:xsecnonprompt}
\end{figure}

In Fig.~\ref{fig:xsecprompt} (left panel) this result is shown as a function of rapidity together with the results from the LHCb experiment at positive (``forward'') and negative (``backward'') rapidity~\cite{Aaij:2013zxa}, corresponding respectively to the p-going and Pb-going direction. 
The \pt-differential cross section of prompt \PJpsi 
is shown, in comparison with ATLAS measurements~\cite{Aaboud:2017cif} at high \pt
and for $-2 < y < 1.5$, in the right panel of Fig.~\ref{fig:xsecprompt}.
The ALICE results, covering the low \pt region at midrapidity, are complementary to the measurements from both the LHCb and ATLAS collaborations.
The data are reported in comparison with 
model calculations 
for prompt \PJpsi (Lansberg et al.~\cite{Lansberg:2016deg,Kusina:2017gkz,Shao:2012iz,Shao:2015vga}) based on the EPPS16~\cite{EPPS16} and the nCTEQ15~\cite{nCTEQ15} sets of nuclear parton distribution functions (nPDFs). 
In both cases, the shaded bands represent the envelope of the computations for different assumptions of the values of the pQCD factorisation ($\mu_{\rm F}$) and renormalisation ($\mu_{\rm R}$) scales (varied within $0.5 < \mu_{\rm F}/\mu_{\rm R} < 2$) computed at the 90\% confidence level.
The predictions show good agreement with data within the large model uncertainties, which are dominated by those on the pQCD scales. 
The results of a Bayesian reweighting approach from the same authors~\cite{Kusina:2017gkz}, employing LHCb measurements of \PJpsi~\cite{Aaij:2017cqq,Aaij:2017gcy} as a constraint for the computations, are also shown in the left panel of Fig.~\ref{fig:xsecprompt}.
Both the size of uncertainties and the difference between the nPDF sets are largely reduced after the reweighting.

In Fig.~\ref{fig:xsecnonprompt}, the cross sections of non-prompt \PJpsi, computed either for $\pt > 0$ (left panel) or differentially in \pt (right panel), are reported together with the corresponding results from the LHCb~\cite{Aaij:2013zxa} and ATLAS~\cite{Aaboud:2017cif} collaborations.
The results are compared with theoretical predictions based on FONLL pQCD calculations~\cite{FONLL} with the inclusion of nuclear shadowing effects according to the  EPPS16 nPDFs~\cite{EPPS16}. 
In each panel, the coloured curves delimit the total theoretical uncertainty on the production cross section, which is dominated by that of the b-quark mass and the pQCD scales, while the shaded bands refer to the theoretical uncertainty of the EPPS16 nPDFs.

\begin{figure}[t]
    \begin{center}
    \includegraphics[width = 0.496\textwidth]{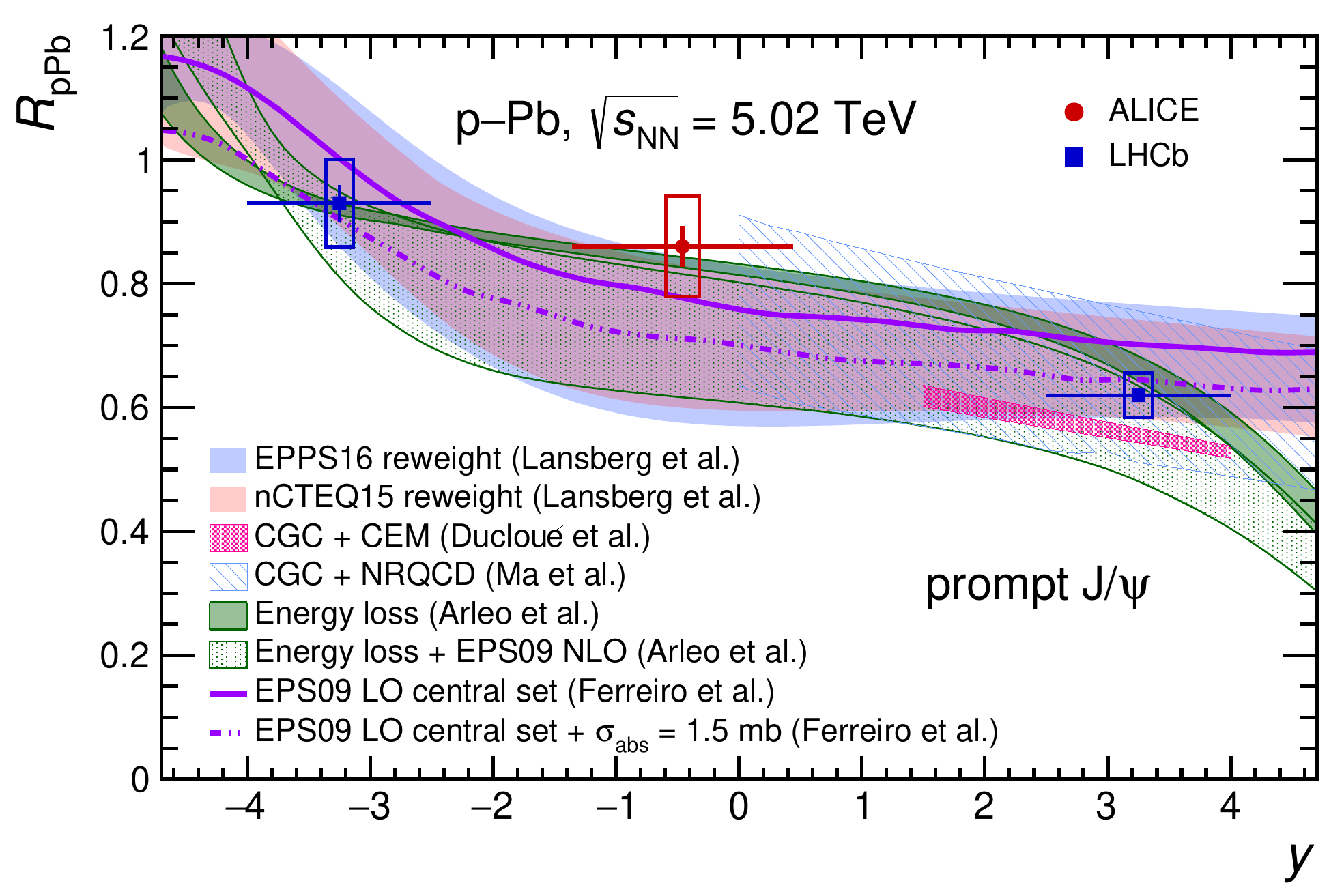}
    \includegraphics[width = 0.496\textwidth]{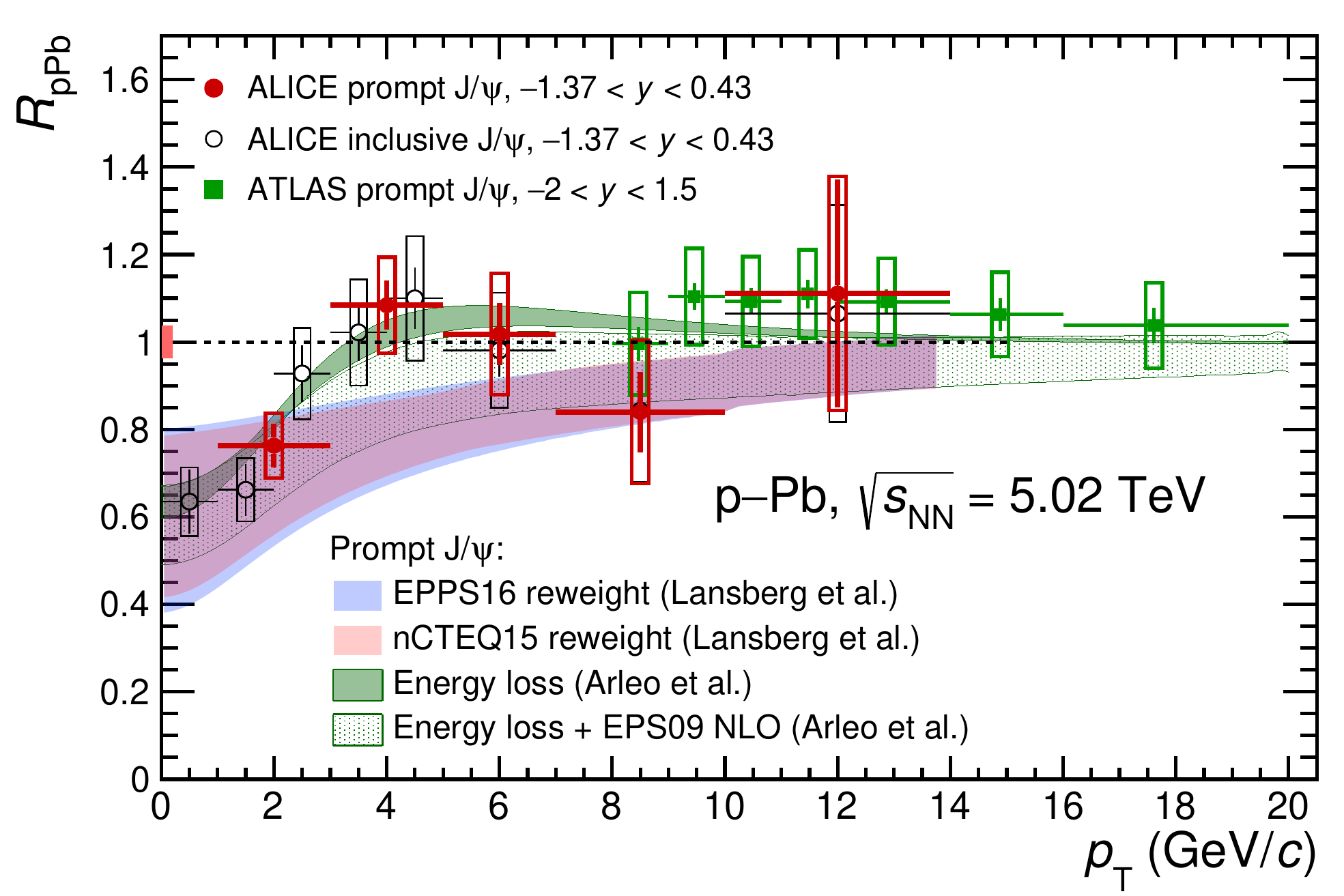} 
    \end{center}
    \caption{$R_{\rm pPb}$ of prompt \PJpsi as a function of rapidity (left panel) and as a function of \pt along with that of inclusive \PJpsi at midrapidity (right panel). 
    Results are shown in comparison with LHCb measurements~\cite{Aaij:2013zxa} at backward and forward rapidity in the left panel and with ATLAS results~\cite{Aaboud:2017cif} (shown up to $\pt = 20$ GeV/$c$) in the right-hand panel.
    Statistical uncertainties are represented by vertical error bars, while open boxes correspond to systematic uncertainties. 
In the left panel, the systematic uncertainty of the ALICE data point includes also the contribution from the extrapolation procedure to go from the visible region ($\pt>1$~GeV/$c$) to $\pt>0$.   
	The filled box around $R_{\rm pPb} = 1$ in the right panel indicates the size of the global relative uncertainty of the ALICE measurements.     
    The results of various model predictions for prompt \PJpsi implementing different CNM effects are also shown~\cite{Kusina:2017gkz,Arleo:2013zua,Ferreiro:2013pua,Ducloue:2015gfa,Ma:2015sia}.}
    \label{fig:RpPbprompt}
\end{figure}

The nuclear modification factor of inclusive \PJpsi, measured for $\pt > 0$ and $-1.37<y<0.43$, amounts to $\rm 0.851 \pm 0.028\ (stat.) \pm 0.079\ (syst.)$. This quantity is obtained using the measured inclusive \PJpsi cross section in pp collisions at $\sqrt{s}=5.02$~TeV~\cite{OurppRefPaper}. Similarly, the \pt-differential $R_{\rm pPb}$ of inclusive \PJpsi is obtained on the basis of the measured pp reference, except for the highest \pt interval (10--14~GeV/$c$), where the statistical sample is limited and the interpolation procedure~\cite{Adam:2015iga} is still used.  
In Fig.~\ref{fig:RpPbprompt}, the $R_{\rm pPb}$ of prompt \PJpsi is reported either for $\pt > 0$ in comparison with LHCb measurements~\cite{Aaij:2013zxa} at backward and forward rapidity (left panel) 
or as a function of \pt, computed according to Eq.~\ref{eqRpPbfB}, together with that of inclusive \PJpsi (right panel) in comparison with ATLAS results~\cite{Aaboud:2017cif}. 
The \pt-integrated $R_{\rm pPb}$ of prompt \PJpsi at midrapidity ($\pt > 0$ and $-1.37<y<0.43$)  is measured to be smaller than unity and amounts to $\rm 0.860 \pm 0.033~(stat.) \pm 0.081~(syst.)$. 
Given also the relatively small fraction of \PJpsi from b-hadron decays for $p_{\rm T} < 14 $~GeV/$c$, the $R_{\rm pPb}$ of inclusive \PJpsi is comparable with that of the prompt component. 
As shown in the right panel of Fig.~\ref{fig:RpPbprompt}, both trends indicate that the suppression observed at midrapidity is a low-\pt effect, concentrated for $\pt \lesssim 3$ GeV/$c$.
The measurements are compared with results from various model predictions which embed different CNM effects {into prompt \PJpsi production}.
In addition to the previously described computations by Lansberg et al., which include a reweighting of the EPPS16 and nCTE15 nPDFs~\cite{Kusina:2017gkz},
the central values of a computation  based on EPS09 nPDF with or without interaction with a nuclear medium (Ferreiro et al.~\cite{Ferreiro:2013pua}) are shown.
A calculation including the effects of coherent energy loss (Arleo et al.~\cite{Arleo:2013zua}), with or without the introduction of nuclear shadowing effects according to EPS09 nPDF, provides a fairly good description of the measurements either as a function of \pt or as a function of rapidity.
Two model calculations based on the CGC effective theory coupled with different  elementary production models (Duclou{\'e} et al.~\cite{Ducloue:2015gfa}, Ma et al.~\cite{Ma:2015sia}), are also reported within their domain of validity, in the forward-$y$ region.

\begin{figure}[h!b]
    \begin{center}
    \includegraphics[width = 0.496\textwidth]{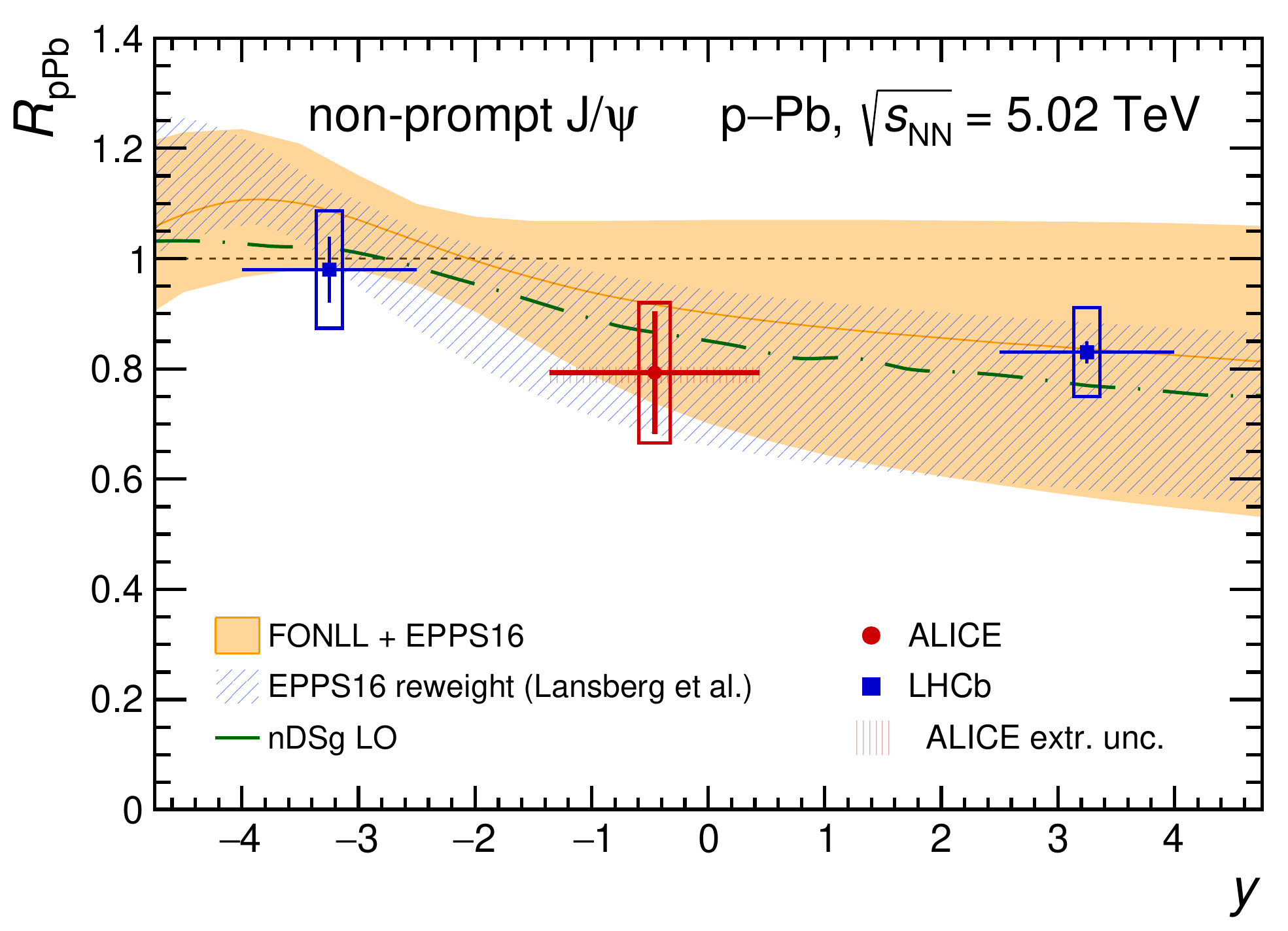}
    \includegraphics[width = 0.496\textwidth]{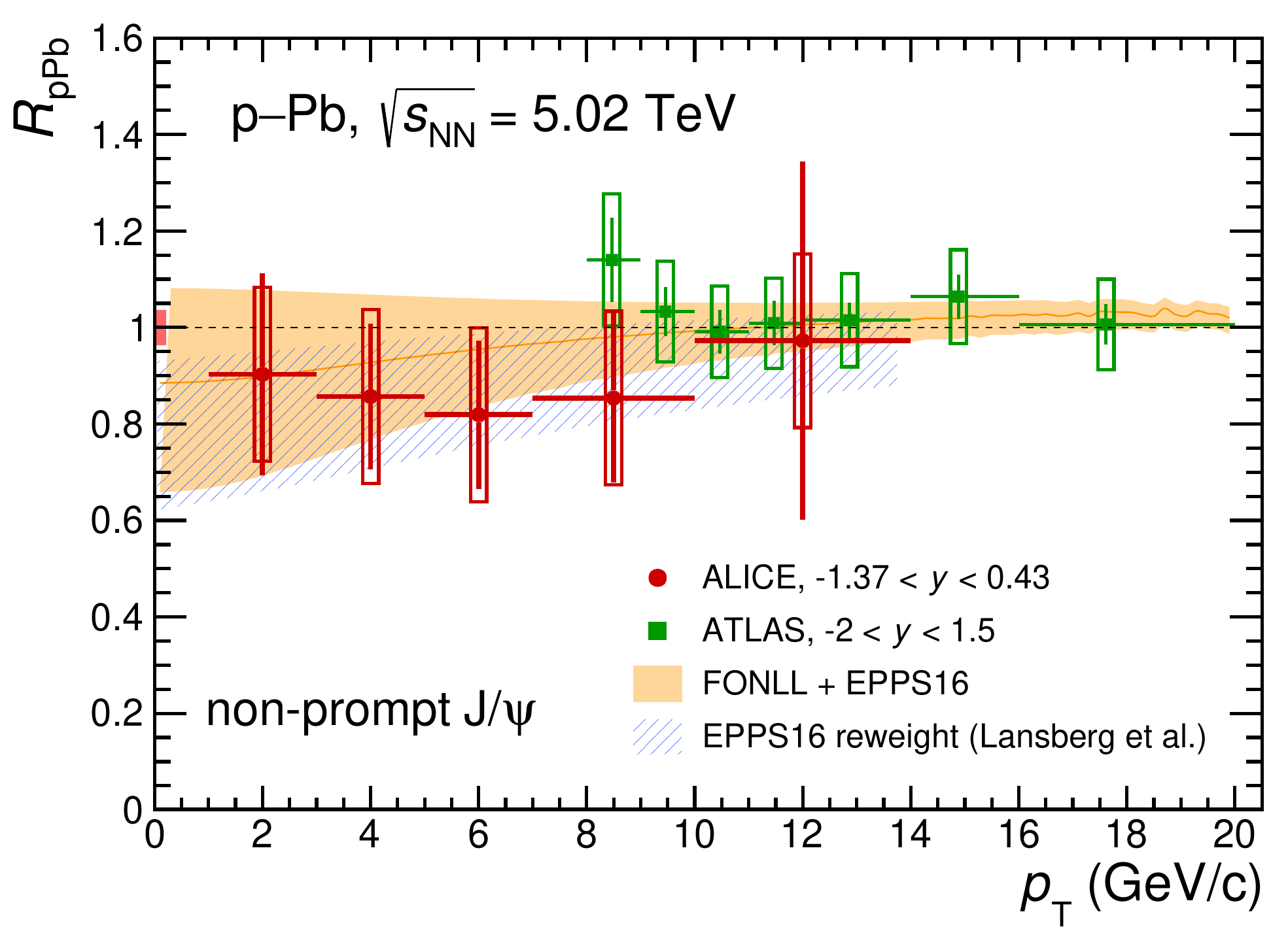} 
    \end{center}
    \caption{Nuclear modification factor $R_{\rm pPb}$ of non-prompt \PJpsi as a function of rapidity (left panel) and as a function of \pt at midrapidity (right panel). The results are compared with similar measurements from the LHCb~\cite{Aaij:2013zxa} and ATLAS~\cite{Aaboud:2017cif} experiments.
    Vertical bars and the open boxes indicate the statistical and systematic uncertainties, respectively. 
    The uncertainty due to the extrapolation to $\pt = 0$ GeV/$c$ for the ALICE measurement is reported as a shaded box in left panel, while the filled box around $R_{\rm pPb} = 1$ in the right-hand panel denotes the size of the global normalisation uncertainty.
    The predicted nuclear modification factors according to the nDSg~\cite{deFlorian:2011fp} (central value shown in the left panel only) and EPPS16~\cite{EPPS16} parameterisations (including a reweighted computation from~\cite{Kusina:2017gkz}) are shown superimposed on both panels.}
    \label{fig:RpPbnonprompt}
\end{figure}

The nuclear modification factor for non-prompt \PJpsi, determined according to Eq.~\ref{eqRpPbfB}, is shown in Fig.~\ref{fig:RpPbnonprompt}.
The measured value of the \pt-integrated $R_{\rm pPb}$ for $\pt > 0$ and $-1.37<y<0.43$ is found to be $\rm 0.79 \pm 0.11\ (stat.) \pm 0.13\ (syst.)\ ^{+0.01}_{-0.02}\ (extr.)$, suggesting the presence of nuclear effects also for the non-prompt \PJpsi component. 
Within uncertainties, the measurements are found to be  compatible with those of the LHCb collaboration ~\cite{Aaij:2013zxa} at both forward and backward rapidity as well as with those of the  ATLAS collaboration~\cite{Aaboud:2017cif} for $\pt \gtrsim 9$ GeV/$c$. 
All data are in fair agreement with the mild degree of suppression predicted by FONLL computations employing the EPPS16 nPDFs.
The nuclear modification factor, as predicted from the Bayesian reweighting approach~\cite{Kusina:2017gkz} of the EPPS16 nPDFs previously introduced for the prompt component, also provides a good description of the measurements.
The central value of an alternative parameterisation of the nuclear PDF, nDSgLO~\cite{deFlorian:2011fp}, is reported in the left-hand panel. 
Despite the larger relative uncertainties, the comparison with the results shown for the prompt component (right panel of Fig.~\ref{fig:RpPbprompt}) suggests that a reduced suppression as well as a less pronounced \pt-dependence affect the component of \PJpsi from b-hadron decays.

Similarly as for our previous work~\cite{OurpPbPaper}, the low \pt coverage of the measured non-prompt \PJpsi cross section at midrapidity, now extending down to $\pt = 1$ GeV/$c$, allows the \pt-integrated b$\rm{\overline{b}}$ cross section per unit of rapidity, $\textup{d}\sigma_{\rm{b\overline{b}}}/\textup{d}y$, and the total b$\rm{\overline{b}}$ production cross section, $\sigma(\rm pPb \rightarrow b\overline{b} + X)$, to be derived with small extrapolation uncertainties. 
By using FONLL with CTEQ6.6 and EPPS16 nPDFs as input model for the computation of the extrapolation factor,
the b$\rm{\overline{b}}$ production cross section at midrapidity is derived as
\begin{linenomath}
\begin{equation}
    \frac{\textup{d}\sigma_{\rm b\overline{b}}}{\textup{d}y}=\frac{\textup{d}\sigma^{\rm model}_{\rm b\overline{b}}}{\textup{d}y}\times\frac{\sigma^{\rm vis}_{\rm \PJpsi\ from\ h_{\rm b}}}{\sigma^{\rm vis,\ model}_{\rm \PJpsi\ from\ h_{\rm b}}}\ .
\end{equation}
\end{linenomath}

Assuming the average branching ratio of \PJpsi from b-hadron decays measured at LEP~\cite{Buskulic:1992wp,Adriani:1993ta,Abreu:1994rk}, $BR(\rm h_b\rightarrow \PJpsi + \textit{X}) = (1.16 \pm 0.10)\%$, for the computation of $\sigma^{\rm vis,\ model}_{\rm \PJpsi\ from\ h_{\rm b}}$,  
the resulting midrapidity cross section per unit of rapidity is
\begin{linenomath}
\begin{equation*}
     \frac{\textup{d}\sigma_{\rm b \overline{b}}}{\textup{d}y} = \rm 5.52 \pm 0.77\ (stat.) \pm 0.75\ (syst.)\ ^{+0.07}_{-0.12}\ (extr.)\ mb.   
\end{equation*}
\end{linenomath}
With a similar approach, the total b$\rm{\overline{b}}$ production cross section is obtained by extrapolating the visible cross section to the full phase space:
\begin{linenomath}
\begin{equation}
    \sigma(\rm p + Pb \rightarrow b\overline{b} + \textit{X}) = \alpha_{\rm 4\pi} \frac{\sigma^{\rm vis}_{\rm \PJpsi\ from\ h_{\rm b}}}{2 \times \textit{BR}(\rm h_b \rightarrow \PJpsi + \textit{X})} ,
\end{equation}
\end{linenomath}
where the factor $\alpha_{\rm 4\pi}$ is defined as the ratio of the yield of non-prompt \PJpsi produced in the full phase space to that in the visible region, and the factor 2 takes into account that 
hadrons with b or $\overline{\rm b}$ valence quark can decay into \PJpsi. 
The value of the $\alpha_{\rm 4\pi}$ factor obtained from FONLL pQCD calculations with EPPS16 nPDFs, with the b-quark fragmentation performed using PYTHIA 6.4 with the Perugia-0 tune, is $\alpha_{\rm 4\pi} = 4.10\ ^{+0.14}_{-0.12}$. 
Using PYTHIA 8 instead of FONLL for the generation of b$\overline{\rm b}$ quark pairs provides
the value 4.02, which is 2\% smaller than the used value for the extrapolation based on FONLL.   
The total cross section is\footnote{The extrapolation uncertainties for the  $\textup{d}\sigma_{\rm{b\overline{b}}}/\textup{d}y$ and for the total b$\rm{\overline{b}}$ cross section include the contributions related to the FONLL, CTEQ6.6 and EPPS16 uncertainties as discussed in Ref.~\cite{OurpPbPaper}, and also a minor contribution of about 0.4\% related to the uncertainty on the \pt-dependence of the relative abundance of b-hadron species.} 
\begin{linenomath}
\begin{equation*}
    \sigma(\rm p + Pb \rightarrow b\overline{b} + \textit{X}) =  \rm 35.5 \pm 5.0\ (stat.) \pm 4.8\ (syst.)\ ^{+1.2}_{-1.0}\ (extr.)\ mb.   
\end{equation*}
\end{linenomath}
The reported results for the extrapolated $\rm b\overline{b}$ production cross sections are consistent with our previous derivations~\cite{OurpPbPaper}. The total uncertainty is  reduced by a factor of 2 thanks to the larger data sample, the smaller systematic uncertainty, and the slightly extended coverage of the visible region where the  non-prompt \PJpsi cross section is measured. 

In p--Pb collisions at $\snn = 5.02$~\TeV the LHCb Collaboration measured the non-prompt \PJpsi\ production cross section 
at forward 
and backward 
rapidities for $\pt<14$~GeV/$c$, reporting~\cite{Aaij:2013zxa}
$\sigma_{\rm \PJpsi\ from\ h_{\rm b}}(1.5<y<4.0)=166.0\pm4.1\pm8.2$~$\mu$b 
and $\sigma_{\rm \PJpsi\ from\ h_{\rm b}}(-5.0<y<-2.5)=118.2\pm6.8\pm11.7$~$\mu$b, respectively. 
A more precise estimate of the total b$\rm{\overline{b}}$ cross section can be obtained by 
repeating the same procedure with  
including also these 
results from the LHCb collaboration~\cite{Aaij:2013zxa}, to obtain a wider visible  
region: $ (-5<y<-2.5, \pt<14~{\rm GeV}/c) \cup  (-1.37<y<0.43, \pt>1.0~{\rm GeV}/c) \cup  (1.5<y<4, \pt<14~{\rm GeV}/c)$. 
The cross section in this wider visible region is obtained as the sum of the cross sections measured in 
this work at central rapidity and those from LHCb. All the uncertainties are uncorrelated except that of the branching ratio. 
In this case, the $\alpha_{\rm 4\pi}$ factor, which is calculated as the ratio of the yield from the model 
in full phase space to that in the wider region covered by the ALICE and LHCb experiments, is reduced to $1.60 \pm 0.02$, and the corresponding total cross section is
\begin{linenomath}
\begin{equation*}
    \sigma(\rm p + Pb \rightarrow b\overline{b} + \textit{X}) =  \rm 33.8 \pm 2.0\ (stat.) \pm 3.4\ (syst.)\ ^{+0.4}_{-0.5}\ (extr.)\ mb\ \ \ \rm (ALICE\ and\ LHCb).   
\end{equation*}
\end{linenomath}

\section{Summary}

The production of \PJpsi mesons in p$-$Pb collisions at $\sqrt{s_{\rm NN}} = 5.02$ TeV is studied based on a data sample about six times larger than that of previously published results
yielding smaller uncertainties and extending the \pt coverage.
The inclusive \PJpsi production cross section at midrapidity is measured down to $\pt = 0$ after reconstructing the \PJpsi mesons in the dielectron decay channel.
The fraction of the inclusive \PJpsi yield originated from b-hadron decays is then determined on a statistical basis, allowing the prompt and non-prompt \PJpsi production cross sections at midrapidity to be derived for $\pt> 1$~GeV/$c$ and as a function of \pt in five momentum intervals. 
The results are scaled to reference measurements from pp collisions at the same centre-of-mass energy in order to investigate the presence of nuclear effects on \PJpsi production.
The nuclear modification factor of prompt \PJpsi shows a significant suppression for $\pt \lesssim 3$ GeV/$c$, whereas there is a hint of 
a less pronounced suppression of the non-prompt component over the inspected \pt range. 
The results can be 
described by
theoretical calculations including various combinations of cold nuclear matter effects, although a precise discrimination among the different models is impaired by the uncertainties affecting the currently available predictions.
Finally, the measurement of the non-prompt \PJpsi production cross section is used to derive the extrapolated midrapidity $\textup{d}\sigma_{\rm b\overline{b}}/\textup{d}y$ and total cross section, $\sigma_{\rm b\overline{b}}$, of beauty quark production.


\newenvironment{acknowledgement}{\relax}{\relax}
\begin{acknowledgement}
\section*{Acknowledgements}

The ALICE Collaboration would like to thank all its engineers and technicians for their invaluable contributions to the construction of the experiment and the CERN accelerator teams for the outstanding performance of the LHC complex.
The ALICE Collaboration gratefully acknowledges the resources and support provided by all Grid centres and the Worldwide LHC Computing Grid (WLCG) collaboration.
The ALICE Collaboration acknowledges the following funding agencies for their support in building and running the ALICE detector:
A. I. Alikhanyan National Science Laboratory (Yerevan Physics Institute) Foundation (ANSL), State Committee of Science and World Federation of Scientists (WFS), Armenia;
Austrian Academy of Sciences, Austrian Science Fund (FWF): [M 2467-N36] and Nationalstiftung f\"{u}r Forschung, Technologie und Entwicklung, Austria;
Ministry of Communications and High Technologies, National Nuclear Research Center, Azerbaijan;
Conselho Nacional de Desenvolvimento Cient\'{\i}fico e Tecnol\'{o}gico (CNPq), Financiadora de Estudos e Projetos (Finep), Funda\c{c}\~{a}o de Amparo \`{a} Pesquisa do Estado de S\~{a}o Paulo (FAPESP) and Universidade Federal do Rio Grande do Sul (UFRGS), Brazil;
Ministry of Education of China (MOEC) , Ministry of Science \& Technology of China (MSTC) and National Natural Science Foundation of China (NSFC), China;
Ministry of Science and Education and Croatian Science Foundation, Croatia;
Centro de Aplicaciones Tecnol\'{o}gicas y Desarrollo Nuclear (CEADEN), Cubaenerg\'{\i}a, Cuba;
Ministry of Education, Youth and Sports of the Czech Republic, Czech Republic;
The Danish Council for Independent Research | Natural Sciences, the VILLUM FONDEN and Danish National Research Foundation (DNRF), Denmark;
Helsinki Institute of Physics (HIP), Finland;
Commissariat \`{a} l'Energie Atomique (CEA) and Institut National de Physique Nucl\'{e}aire et de Physique des Particules (IN2P3) and Centre National de la Recherche Scientifique (CNRS), France;
Bundesministerium f\"{u}r Bildung und Forschung (BMBF) and GSI Helmholtzzentrum f\"{u}r Schwerionenforschung GmbH, Germany;
General Secretariat for Research and Technology, Ministry of Education, Research and Religions, Greece;
National Research, Development and Innovation Office, Hungary;
Department of Atomic Energy Government of India (DAE), Department of Science and Technology, Government of India (DST), University Grants Commission, Government of India (UGC) and Council of Scientific and Industrial Research (CSIR), India;
Indonesian Institute of Science, Indonesia;
Istituto Nazionale di Fisica Nucleare (INFN), Italy;
Institute for Innovative Science and Technology , Nagasaki Institute of Applied Science (IIST), Japanese Ministry of Education, Culture, Sports, Science and Technology (MEXT) and Japan Society for the Promotion of Science (JSPS) KAKENHI, Japan;
Consejo Nacional de Ciencia (CONACYT) y Tecnolog\'{i}a, through Fondo de Cooperaci\'{o}n Internacional en Ciencia y Tecnolog\'{i}a (FONCICYT) and Direcci\'{o}n General de Asuntos del Personal Academico (DGAPA), Mexico;
Nederlandse Organisatie voor Wetenschappelijk Onderzoek (NWO), Netherlands;
The Research Council of Norway, Norway;
Commission on Science and Technology for Sustainable Development in the South (COMSATS), Pakistan;
Pontificia Universidad Cat\'{o}lica del Per\'{u}, Peru;
Ministry of Education and Science, National Science Centre and WUT ID-UB, Poland;
Korea Institute of Science and Technology Information and National Research Foundation of Korea (NRF), Republic of Korea;
Ministry of Education and Scientific Research, Institute of Atomic Physics and Ministry of Research and Innovation and Institute of Atomic Physics, Romania;
Joint Institute for Nuclear Research (JINR), Ministry of Education and Science of the Russian Federation, National Research Centre Kurchatov Institute, Russian Science Foundation and Russian Foundation for Basic Research, Russia;
Ministry of Education, Science, Research and Sport of the Slovak Republic, Slovakia;
National Research Foundation of South Africa, South Africa;
Swedish Research Council (VR) and Knut \& Alice Wallenberg Foundation (KAW), Sweden;
European Organization for Nuclear Research, Switzerland;
Suranaree University of Technology (SUT), National Science and Technology Development Agency (NSDTA) and Office of the Higher Education Commission under NRU project of Thailand, Thailand;
Turkish Energy, Nuclear and Mineral Research Agency (TENMAK), Turkey;
National Academy of  Sciences of Ukraine, Ukraine;
Science and Technology Facilities Council (STFC), United Kingdom;
National Science Foundation of the United States of America (NSF) and United States Department of Energy, Office of Nuclear Physics (DOE NP), United States of America.

In addition, individual groups and members have received support from the Horizon 2020 programme, European Union.

\end{acknowledgement}

\bibliographystyle{utphys}   
\bibliography{bibliography}

\newpage
\appendix

%
%

\section{The ALICE Collaboration}
\label{app:collab}
%
\begingroup
\small
\begin{flushleft}
S.~Acharya$^{\rm 143}$, 
D.~Adamov\'{a}$^{\rm 98}$, 
A.~Adler$^{\rm 76}$, 
J.~Adolfsson$^{\rm 83}$, 
G.~Aglieri Rinella$^{\rm 35}$, 
M.~Agnello$^{\rm 31}$, 
N.~Agrawal$^{\rm 55}$, 
Z.~Ahammed$^{\rm 143}$, 
S.~Ahmad$^{\rm 16}$, 
S.U.~Ahn$^{\rm 78}$
% ...

\section*{Affiliation notes}

$^{\rm I}$ Deceased\\
$^{\rm II}$ Also at: Italian National Agency for New Technologies, Energy and Sustainable Economic Development (ENEA), Bologna, Italy\\
$^{\rm III}$ Also at: Dipartimento DET del Politecnico di Torino, Turin, Italy\\
$^{\rm IV}$ Also at: M.V. Lomonosov Moscow State University, D.V. Skobeltsyn Institute of Nuclear, Physics, Moscow, Russia\\
$^{\rm V}$ Also at: Institute of Theoretical Physics, University of Wroclaw, Poland\\

\section*{Collaboration Institutes}

$^{1}$ A.I. Alikhanyan National Science Laboratory (Yerevan Physics Institute) Foundation, Yerevan, Armenia\\
$^{2}$ AGH University of Science and Technology, Cracow, Poland\\
$^{3}$ Bogolyubov Institute for Theoretical Physics, National Academy of Sciences of Ukraine, Kiev, Ukraine\\
$^{4}$ Bose Institute, Department of Physics  and Centre for Astroparticle Physics and Space Science (CAPSS), Kolkata, India\\
$^{5}$ Budker Institute for Nuclear Physics, Novosibirsk, Russia\\
$^{6}$ California Polytechnic State University, San Luis Obispo, California, United States\\
$^{7}$ Central China Normal University, Wuhan, China\\
$^{8}$ Centro de Aplicaciones Tecnol\'{o}gicas y Desarrollo Nuclear (CEADEN), Havana, Cuba\\
$^{9}$ Centro de Investigaci\'{o}n y de Estudios Avanzados (CINVESTAV), Mexico City and M\'{e}rida, Mexico\\
$^{10}$ Chicago State University, Chicago, Illinois, United States\\
$^{11}$ China Institute of Atomic Energy, Beijing, China\\
$^{12}$ Chungbuk National University, Cheongju, Republic of Korea\\
$^{13}$ Comenius University Bratislava, Faculty of Mathematics, Physics and Informatics, Bratislava, Slovakia\\
$^{14}$ COMSATS University Islamabad, Islamabad, Pakistan\\
$^{15}$ Creighton University, Omaha, Nebraska, United States\\
$^{16}$ Department of Physics, Aligarh Muslim University, Aligarh, India\\
$^{17}$ Department of Physics, Pusan National University, Pusan, Republic of Korea\\
$^{18}$ Department of Physics, Sejong University, Seoul, Republic of Korea\\
$^{19}$ Department of Physics, University of California, Berkeley, California, United States\\
$^{20}$ Department of Physics, University of Oslo, Oslo, Norway\\
$^{21}$ Department of Physics and Technology, University of Bergen, Bergen, Norway\\
$^{22}$ Dipartimento di Fisica dell'Universit\`{a} 'La Sapienza' and Sezione INFN, Rome, Italy\\
$^{23}$ Dipartimento di Fisica dell'Universit\`{a} and Sezione INFN, Cagliari, Italy\\
$^{24}$ Dipartimento di Fisica dell'Universit\`{a} and Sezione INFN, Trieste, Italy\\
$^{25}$ Dipartimento di Fisica dell'Universit\`{a} and Sezione INFN, Turin, Italy\\
$^{26}$ Dipartimento di Fisica e Astronomia dell'Universit\`{a} and Sezione INFN, Bologna, Italy\\
$^{27}$ Dipartimento di Fisica e Astronomia dell'Universit\`{a} and Sezione INFN, Catania, Italy\\
$^{28}$ Dipartimento di Fisica e Astronomia dell'Universit\`{a} and Sezione INFN, Padova, Italy\\
$^{29}$ Dipartimento di Fisica e Nucleare e Teorica, Universit\`{a} di Pavia, Pavia, Italy\\
$^{30}$ Dipartimento di Fisica `E.R.~Caianiello' dell'Universit\`{a} and Gruppo Collegato INFN, Salerno, Italy\\
$^{31}$ Dipartimento DISAT del Politecnico and Sezione INFN, Turin, Italy\\
$^{32}$ Dipartimento di Scienze e Innovazione Tecnologica dell'Universit\`{a} del Piemonte Orientale and INFN Sezione di Torino, Alessandria, Italy\\
$^{33}$ Dipartimento di Scienze MIFT, Universit\`{a} di Messina, Messina, Italy\\
$^{34}$ Dipartimento Interateneo di Fisica `M.~Merlin' and Sezione INFN, Bari, Italy\\
$^{35}$ European Organization for Nuclear Research (CERN), Geneva, Switzerland\\
$^{36}$ Faculty of Electrical Engineering, Mechanical Engineering and Naval Architecture, University of Split, Split, Croatia\\
$^{37}$ Faculty of Engineering and Science, Western Norway University of Applied Sciences, Bergen, Norway\\
$^{38}$ Faculty of Nuclear Sciences and Physical Engineering, Czech Technical University in Prague, Prague, Czech Republic\\
$^{39}$ Faculty of Science, P.J.~\v{S}af\'{a}rik University, Ko\v{s}ice, Slovakia\\
$^{40}$ Frankfurt Institute for Advanced Studies, Johann Wolfgang Goethe-Universit\"{a}t Frankfurt, Frankfurt, Germany\\
$^{41}$ Fudan University, Shanghai, China\\
$^{42}$ Gangneung-Wonju National University, Gangneung, Republic of Korea\\
$^{43}$ Gauhati University, Department of Physics, Guwahati, India\\
$^{44}$ Helmholtz-Institut f\"{u}r Strahlen- und Kernphysik, Rheinische Friedrich-Wilhelms-Universit\"{a}t Bonn, Bonn, Germany\\
$^{45}$ Helsinki Institute of Physics (HIP), Helsinki, Finland\\
$^{46}$ High Energy Physics Group,  Universidad Aut\'{o}noma de Puebla, Puebla, Mexico\\
$^{47}$ Hiroshima University, Hiroshima, Japan\\
$^{48}$ Hochschule Worms, Zentrum  f\"{u}r Technologietransfer und Telekommunikation (ZTT), Worms, Germany\\
$^{49}$ Horia Hulubei National Institute of Physics and Nuclear Engineering, Bucharest, Romania\\
$^{50}$ Indian Institute of Technology Bombay (IIT), Mumbai, India\\
$^{51}$ Indian Institute of Technology Indore, Indore, India\\
$^{52}$ Indonesian Institute of Sciences, Jakarta, Indonesia\\
$^{53}$ INFN, Laboratori Nazionali di Frascati, Frascati, Italy\\
$^{54}$ INFN, Sezione di Bari, Bari, Italy\\
$^{55}$ INFN, Sezione di Bologna, Bologna, Italy\\
$^{56}$ INFN, Sezione di Cagliari, Cagliari, Italy\\
$^{57}$ INFN, Sezione di Catania, Catania, Italy\\
$^{58}$ INFN, Sezione di Padova, Padova, Italy\\
$^{59}$ INFN, Sezione di Pavia, Pavia, Italy\\
$^{60}$ INFN, Sezione di Roma, Rome, Italy\\
$^{61}$ INFN, Sezione di Torino, Turin, Italy\\
$^{62}$ INFN, Sezione di Trieste, Trieste, Italy\\
$^{63}$ Inha University, Incheon, Republic of Korea\\
$^{64}$ Institute for Gravitational and Subatomic Physics (GRASP), Utrecht University/Nikhef, Utrecht, Netherlands\\
$^{65}$ Institute for Nuclear Research, Academy of Sciences, Moscow, Russia\\
$^{66}$ Institute of Experimental Physics, Slovak Academy of Sciences, Ko\v{s}ice, Slovakia\\
$^{67}$ Institute of Physics, Homi Bhabha National Institute, Bhubaneswar, India\\
$^{68}$ Institute of Physics of the Czech Academy of Sciences, Prague, Czech Republic\\
$^{69}$ Institute of Space Science (ISS), Bucharest, Romania\\
$^{70}$ Institut f\"{u}r Kernphysik, Johann Wolfgang Goethe-Universit\"{a}t Frankfurt, Frankfurt, Germany\\
$^{71}$ Instituto de Ciencias Nucleares, Universidad Nacional Aut\'{o}noma de M\'{e}xico, Mexico City, Mexico\\
$^{72}$ Instituto de F\'{i}sica, Universidade Federal do Rio Grande do Sul (UFRGS), Porto Alegre, Brazil\\
$^{73}$ Instituto de F\'{\i}sica, Universidad Nacional Aut\'{o}noma de M\'{e}xico, Mexico City, Mexico\\
$^{74}$ iThemba LABS, National Research Foundation, Somerset West, South Africa\\
$^{75}$ Jeonbuk National University, Jeonju, Republic of Korea\\
$^{76}$ Johann-Wolfgang-Goethe Universit\"{a}t Frankfurt Institut f\"{u}r Informatik, Fachbereich Informatik und Mathematik, Frankfurt, Germany\\
$^{77}$ Joint Institute for Nuclear Research (JINR), Dubna, Russia\\
$^{78}$ Korea Institute of Science and Technology Information, Daejeon, Republic of Korea\\
$^{79}$ KTO Karatay University, Konya, Turkey\\
$^{80}$ Laboratoire de Physique des 2 Infinis, Ir\`{e}ne Joliot-Curie, Orsay, France\\
$^{81}$ Laboratoire de Physique Subatomique et de Cosmologie, Universit\'{e} Grenoble-Alpes, CNRS-IN2P3, Grenoble, France\\
$^{82}$ Lawrence Berkeley National Laboratory, Berkeley, California, United States\\
$^{83}$ Lund University Department of Physics, Division of Particle Physics, Lund, Sweden\\
$^{84}$ Moscow Institute for Physics and Technology, Moscow, Russia\\
$^{85}$ Nagasaki Institute of Applied Science, Nagasaki, Japan\\
$^{86}$ Nara Women{'}s University (NWU), Nara, Japan\\
$^{87}$ National and Kapodistrian University of Athens, School of Science, Department of Physics , Athens, Greece\\
$^{88}$ National Centre for Nuclear Research, Warsaw, Poland\\
$^{89}$ National Institute of Science Education and Research, Homi Bhabha National Institute, Jatni, India\\
$^{90}$ National Nuclear Research Center, Baku, Azerbaijan\\
$^{91}$ National Research Centre Kurchatov Institute, Moscow, Russia\\
$^{92}$ Niels Bohr Institute, University of Copenhagen, Copenhagen, Denmark\\
$^{93}$ Nikhef, National institute for subatomic physics, Amsterdam, Netherlands\\
$^{94}$ NRC Kurchatov Institute IHEP, Protvino, Russia\\
$^{95}$ NRC \guillemotleft Kurchatov\guillemotright  Institute - ITEP, Moscow, Russia\\
$^{96}$ NRNU Moscow Engineering Physics Institute, Moscow, Russia\\
$^{97}$ Nuclear Physics Group, STFC Daresbury Laboratory, Daresbury, United Kingdom\\
$^{98}$ Nuclear Physics Institute of the Czech Academy of Sciences, \v{R}e\v{z} u Prahy, Czech Republic\\
$^{99}$ Oak Ridge National Laboratory, Oak Ridge, Tennessee, United States\\
$^{100}$ Ohio State University, Columbus, Ohio, United States\\
$^{101}$ Petersburg Nuclear Physics Institute, Gatchina, Russia\\
$^{102}$ Physics department, Faculty of science, University of Zagreb, Zagreb, Croatia\\
$^{103}$ Physics Department, Panjab University, Chandigarh, India\\
$^{104}$ Physics Department, University of Jammu, Jammu, India\\
$^{105}$ Physics Department, University of Rajasthan, Jaipur, India\\
$^{106}$ Physikalisches Institut, Eberhard-Karls-Universit\"{a}t T\"{u}bingen, T\"{u}bingen, Germany\\
$^{107}$ Physikalisches Institut, Ruprecht-Karls-Universit\"{a}t Heidelberg, Heidelberg, Germany\\
$^{108}$ Physik Department, Technische Universit\"{a}t M\"{u}nchen, Munich, Germany\\
$^{109}$ Politecnico di Bari and Sezione INFN, Bari, Italy\\
$^{110}$ Research Division and ExtreMe Matter Institute EMMI, GSI Helmholtzzentrum f\"ur Schwerionenforschung GmbH, Darmstadt, Germany\\
$^{111}$ Russian Federal Nuclear Center (VNIIEF), Sarov, Russia\\
$^{112}$ Saha Institute of Nuclear Physics, Homi Bhabha National Institute, Kolkata, India\\
$^{113}$ School of Physics and Astronomy, University of Birmingham, Birmingham, United Kingdom\\
$^{114}$ Secci\'{o}n F\'{\i}sica, Departamento de Ciencias, Pontificia Universidad Cat\'{o}lica del Per\'{u}, Lima, Peru\\
$^{115}$ St. Petersburg State University, St. Petersburg, Russia\\
$^{116}$ Stefan Meyer Institut f\"{u}r Subatomare Physik (SMI), Vienna, Austria\\
$^{117}$ SUBATECH, IMT Atlantique, Universit\'{e} de Nantes, CNRS-IN2P3, Nantes, France\\
$^{118}$ Suranaree University of Technology, Nakhon Ratchasima, Thailand\\
$^{119}$ Technical University of Ko\v{s}ice, Ko\v{s}ice, Slovakia\\
$^{120}$ The Henryk Niewodniczanski Institute of Nuclear Physics, Polish Academy of Sciences, Cracow, Poland\\
$^{121}$ The University of Texas at Austin, Austin, Texas, United States\\
$^{122}$ Universidad Aut\'{o}noma de Sinaloa, Culiac\'{a}n, Mexico\\
$^{123}$ Universidade de S\~{a}o Paulo (USP), S\~{a}o Paulo, Brazil\\
$^{124}$ Universidade Estadual de Campinas (UNICAMP), Campinas, Brazil\\
$^{125}$ Universidade Federal do ABC, Santo Andre, Brazil\\
$^{126}$ University of Cape Town, Cape Town, South Africa\\
$^{127}$ University of Houston, Houston, Texas, United States\\
$^{128}$ University of Jyv\"{a}skyl\"{a}, Jyv\"{a}skyl\"{a}, Finland\\
$^{129}$ University of Kansas, Lawrence, Kansas, United States\\
$^{130}$ University of Liverpool, Liverpool, United Kingdom\\
$^{131}$ University of Science and Technology of China, Hefei, China\\
$^{132}$ University of South-Eastern Norway, Tonsberg, Norway\\
$^{133}$ University of Tennessee, Knoxville, Tennessee, United States\\
$^{134}$ University of the Witwatersrand, Johannesburg, South Africa\\
$^{135}$ University of Tokyo, Tokyo, Japan\\
$^{136}$ University of Tsukuba, Tsukuba, Japan\\
$^{137}$ Universit\'{e} Clermont Auvergne, CNRS/IN2P3, LPC, Clermont-Ferrand, France\\
$^{138}$ Universit\'{e} de Lyon, CNRS/IN2P3, Institut de Physique des 2 Infinis de Lyon , Lyon, France\\
$^{139}$ Universit\'{e} de Strasbourg, CNRS, IPHC UMR 7178, F-67000 Strasbourg, France, Strasbourg, France\\
$^{140}$ Universit\'{e} Paris-Saclay Centre d'Etudes de Saclay (CEA), IRFU, D\'{e}partment de Physique Nucl\'{e}aire (DPhN), Saclay, France\\
$^{141}$ Universit\`{a} degli Studi di Foggia, Foggia, Italy\\
$^{142}$ Universit\`{a} di Brescia, Brescia, Italy\\
$^{143}$ Variable Energy Cyclotron Centre, Homi Bhabha National Institute, Kolkata, India\\
$^{144}$ Warsaw University of Technology, Warsaw, Poland\\
$^{145}$ Wayne State University, Detroit, Michigan, United States\\
$^{146}$ Westf\"{a}lische Wilhelms-Universit\"{a}t M\"{u}nster, Institut f\"{u}r Kernphysik, M\"{u}nster, Germany\\
$^{147}$ Wigner Research Centre for Physics, Budapest, Hungary\\
$^{148}$ Yale University, New Haven, Connecticut, United States\\
$^{149}$ Yonsei University, Seoul, Republic of Korea\\

\bigskip 

\end{flushleft} 
\endgroup
  
\end{document}